\documentclass[12pt]{article}

\usepackage{caption}
\usepackage{subcaption}
\usepackage{authblk}

\usepackage{amsmath}
\usepackage{amsfonts}
\usepackage{url}
\usepackage{amsthm}
\usepackage{color}
\usepackage{graphicx}
\usepackage{float}

\usepackage[margin=1in]{geometry}
\begin{document}
\title{Optimal fenestration of the Fontan circulation }
\author[1]{Zan Ahmad}
\author[1,2]{Lynn H. Jin}
\author[3]{Daniel J. Penny}
\author[3]{\\Craig G. Rusin}
\author[1]{Charles S. Peskin} 
\author[1,3]{Charles Puelz}
\affil[1]{Courant Institute of Mathematical Sciences, New York University}
\affil[2]{School of Physics, Georgia Institute of Technology}
\affil[3]{Department of Pediatrics, Section of Cardiology, Baylor College of Medicine and Texas Children's Hospital}
\maketitle

\begin{abstract}
  In this paper, we develop a pulsatile compartmental model of the Fontan circulation and use it to explore the effects of a fenestration added to this physiology. A fenestration is a shunt between the systemic and pulmonary veins that is added either at the time of Fontan conversion or at a later time for the treatment of complications. This shunt increases cardiac output and decreases systemic venous pressure.  However, these hemodynamic benefits are achieved at the expense of a decrease in the arterial oxygen saturation. The model developed this paper incorporates fenestration size as a parameter and describes both blood flow and oxygen transport. It is calibrated to clinical data from Fontan patients, and we use it to study the impact of a fenestration on several hemodynamic variables. In certain scenarios corresponding to high-risk Fontan physiology, we demonstrate the existence of an optimal fenestration size that maximizes oxygen delivery to the systemic tissues. 
\end{abstract}
\section{Introduction}

\begin{figure}[h!]
\begin{center}
  \includegraphics[scale=0.6,trim=0 0 -30 0]{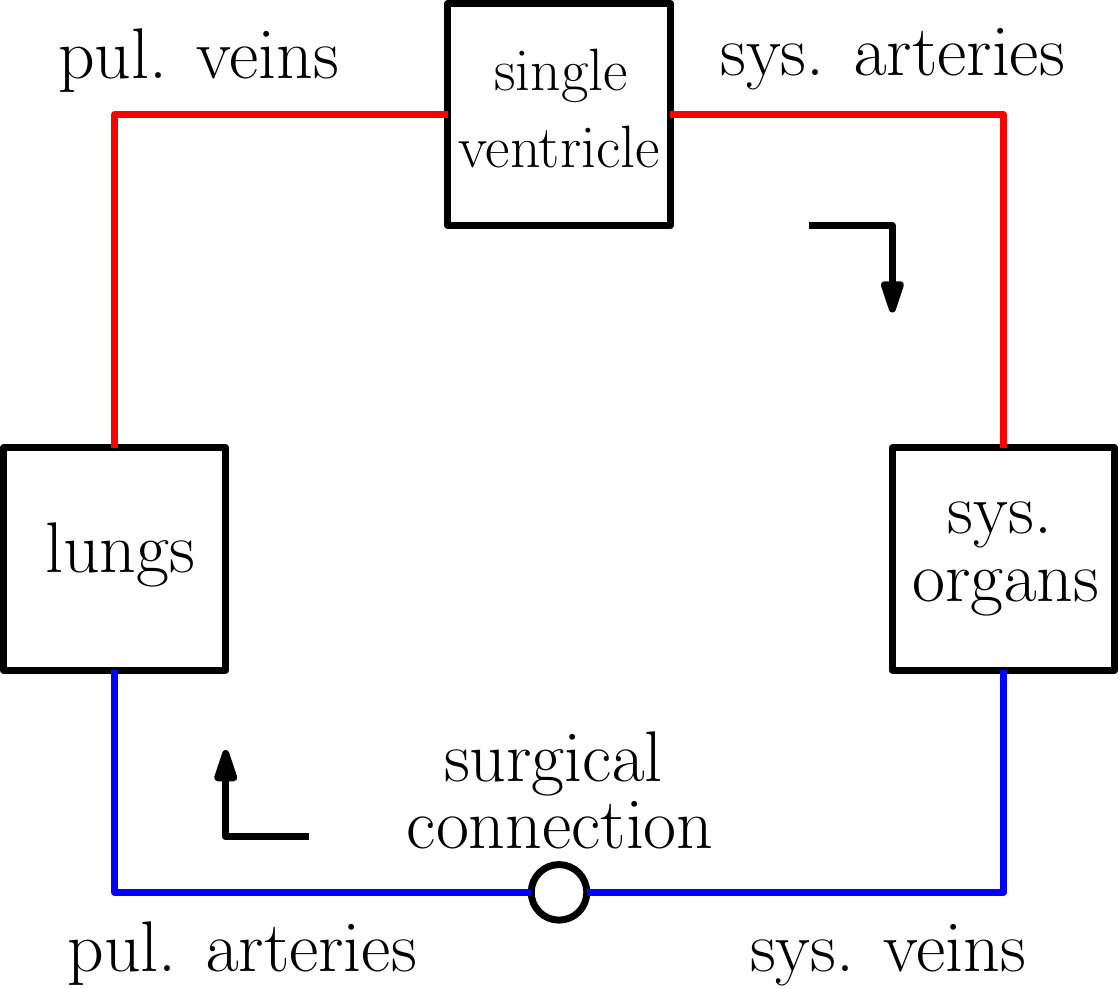}
  \includegraphics[scale=0.6]{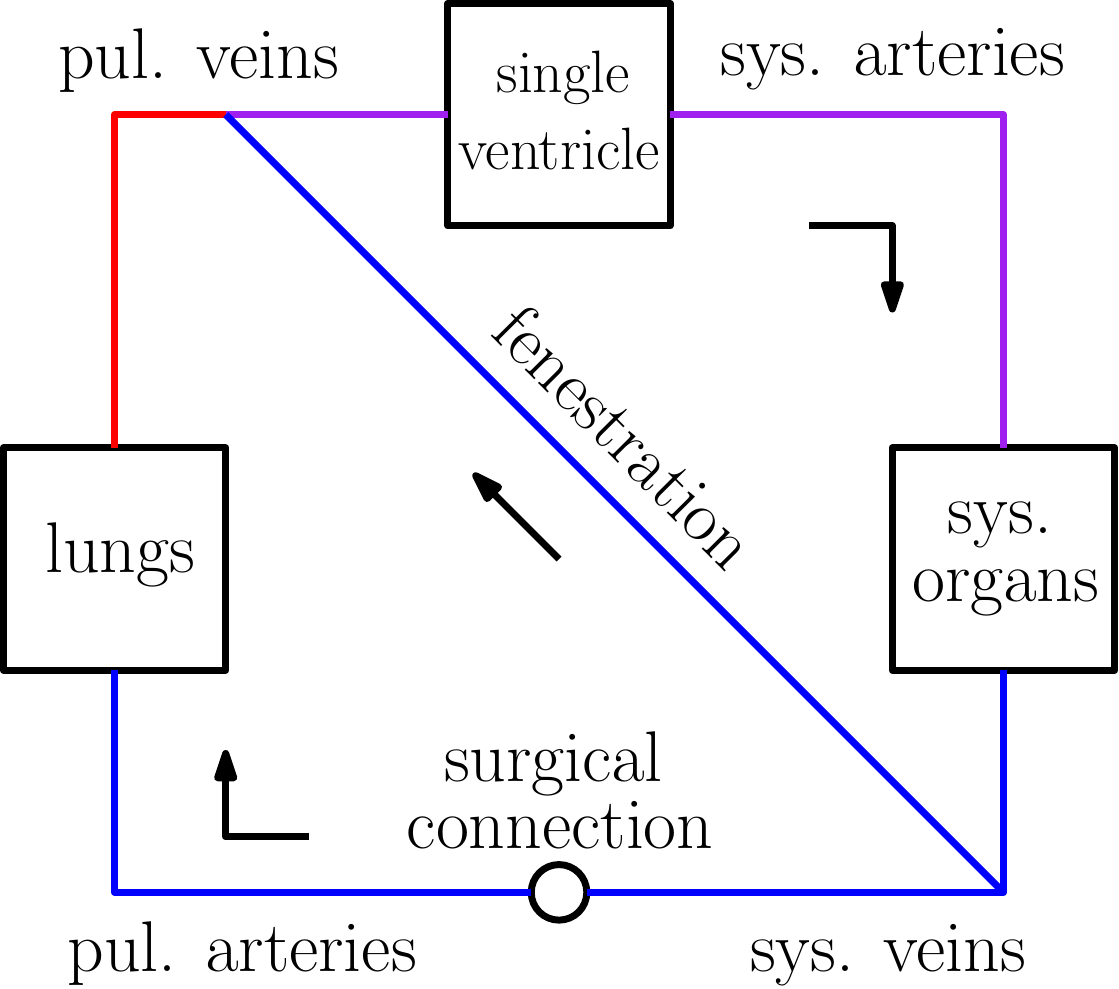}
\caption{Sketches of the Fontan circulation on the left and the fenestrated Fontan circulation on the right. Blue represents deoxygenated blood, red represents oxygenated blood, and purple represents a mixture of oxygenated and deoxygenated blood.}
\label{fig:Fontan}
  \end{center}
\end{figure}

 Single ventricle physiology corresponds to a spectrum of congenital heart defects in which there is only one functioning ventricular chamber. Patients with this type of condition require complex medical and surgical interventions to ensure survival. The typical course of treatment for these defects is a sequence of surgeries during the first several years of life, ending with a procedure that establishes an abnormal physiology known as the Fontan circulation. This physiology was conceived in 1971 and is characterized by the systemic organs and lungs in series, as in a normal circulation \cite{Fontan71}. However, in contrast to a normal circulation, the Fontan physiology relies on passive blood flow to the lungs. This circulation is created by surgically placing the single functioning ventricle upstream from the systemic organs and connecting the vena cavae directly to the pulmonary arteries. Refer to the left panel of Figure (\ref{fig:Fontan}) for a schematic of the Fontan circulation, where the label ``surgical connection'' corresponds to the connection established between the vena cavae and pulmonary arteries. It is important to note that this connection has relatively low resistance, resulting in a negligible difference between the systemic venous and pulmonary artery pressures.

Fontan patients may experience severe complications because of their abnormal physiology, including protein-losing enteropathy, ventricular and hepatic dysfunction, and plastic bronchitis \cite{Feldt96, Mondesert13, Barracano22}. Many of these issues might be attributed to chronically low cardiac output because of the serialized organs and lungs leading to higher than normal systemic venous pressure \cite{Rychik12}. More specifically, the low resistance in the surgical connection between the pulmonary arteries and the vena cavae results in a pulmonary artery pressure that approximately equals the systemic venous pressure. In turn, this might lead to an elevated systemic venous pressure compared to that found in a normal circulation. Higher systemic venous pressure is thought to be a cause of complications \cite{Rychik12, Ohuchi17}. Theoretically, cardiac output can be increased while pulmonary artery and systemic venous pressures can be simultaneously decreased through the introduction of a shunt between the systemic veins and the pulmonary veins known as a fenestration. This additional conduit has been introduced into the Fontan physiology both at the time of Fontan conversion and also later in the patient's life for the treatment of complications \cite{Lemler02, Rychik97}. While a fenestration typically increases cardiac output, it also decreases oxygen saturation of blood flowing to the systemic tissues. This is because blood flowing through the fenestration bypasses the lungs and is therefore not re-oxygenated. For a schematic of the Fontan circulation with a fenestration, refer to the right panel of Figure \ref{fig:Fontan}. The trade-off between a decrease in oxygen saturation and an increase in cardiac output can also be seen in the Norwood physiology, which was mathematically studied using a non-pulsatile model by Barnea et al. \cite{Barnea94}. The authors demonstrated that the Norwood circulation can be ``balanced'' to optimize oxygen delivery to the systemic tissues. This paper is concerned with similar questions regarding the Fontan physiology. We believe this study, focused on fenestrating the Fontan circulation, is of clinical relevance because of the aging Fontan population and possible therapeutic benefits of such an intervention \cite{Barracano22, Rychik97}.

The complexity of the Fontan circulation has motivated the development of many mathematical models which focus on various aspects of this physiology. For an extensive review of modeling efforts devoted to the Fontan physiology, see \cite{Degroff08}. Researchers have created three dimensional fluid dynamics models for the connection between the vena cavae and pulmonary arteries, called the total cavopulmonary connection in an extracardiac Fontan circulation, in an effort to minimize energy loss in the surgical anatomy \cite{Marsden09, Yang10}. There has also been work using lower dimensional models that include one-dimensional or compartmental descriptions for fluid mechanics. Puelz et al. developed a one-dimensional/compartmental model of the Fontan physiology with a fenestration that was combined with a non-pulsatile description of oxygen transport \cite{Puelz17}. Conover et al. used compartmental modeling to study the surgical stages for hypoplastic left heart syndrome, including the Fontan physiology \cite{Conover18}. Liang et al. constructed compartmental models for both normal and Fontan circulations in order to systematically compare their hemodynamics \cite{Liang14}. To our knowledge, the present paper describes the first model of the fenestrated Fontan circulation that employs a fully time-dependent description for both hemodynamics and oxygen transport.

We develop a pulsatile compartmental model of the Fontan physiology and use it to study the impact of a fenestration on blood flow and oxygen transport. The model is calibrated to clinical data with the fenestration closed in order to provide a baseline set of parameters. We vary both the pulmonary vascular resistance and fenestration size, with respect to the baseline model, to study the fenestration's impact on several important variables, including systemic venous pressure, cardiac output, oxygen saturation, and oxygen delivery. Our model demonstrates that in certain scenarios which correspond to high-risk Fontan physiology, a fenestration size exists which optimizes oxygen delivery while simultaneously decreasing systemic venous pressure. Although the predicted increase in oxygen delivery is modest, the fact that there is an increase shows at least that there is no detriment in oxygen delivery through the introduction of an optimally-sized fenestration. Our model also predicts that such a fenestration decreases the systemic venous pressure, an effect that could be of substantial benefit to the patient.

\section{Models and methods}
\label{section:mathmodels}

Our model for the Fontan circulation contains two main parts: (i) a blood flow model that incorporates a nonlinear resistance for the fenestration and (ii) an oxygen transport model. The blood flow model is presented in the first subsection along with details for modeling the fenestration. The oxygen transport model is described in the second subsection. Methods for numerically approximating these models are detailed in the final subsection. The models and methods presented here closely follow those given in related work by Han et al., and additional details may be found in that paper \cite{Han21}.

\subsection{Blood flow model}
\label{section:pulsatile}
 
 The blood flow model used here describes the circulation as a series of compartments that are either compliance chambers or resistor elements.  Our approach is derived from earlier work by Peskin and Tu, with modifications to account for the Fontan physiology and fenestration \cite{Hoppensteadt13, Peskin86, Tu89}.  Each chamber is identified with a unique index $i$. We assume the following relation between compliance $C_i$, pressure $P_i$, and volume $V_i$:
\begin{align}
V_i = V_{i,\text{d}} + C_i P_i.
\label{eq:comp_chamber}
\end{align}
The dead volume $V_{i,\text{d}}$ is equal to the residual chamber volume at zero pressure. The heart chambers obey equation \eqref{eq:comp_chamber}, but in this case the compliance is taken to be a time-dependent function that varies periodically between minimum $C_\text{systole}$ and maximum $C_\text{diastole}$. Since our model describes single ventricle physiology, one side of the heart is removed from the circulation. The single ventricle in the Fontan circulation receives blood from the pulmonary veins and ejects that blood into the systemic arteries (refer to Figure \ref{fig:Fontan}). The single ventricle is a distinct chamber with a time-varying compliance and the atrium chambers are lumped in with the pulmonary venous chamber.

The time-dependent ventricular compliance, denoted $C_\text{ventricle} = C_\text{ventricle}(t)$, is taken to be a periodic function of time, with period $T$ corresponding to the duration of the cardiac cycle. The compliance is defined as the reciprocal of the elastance $E_\text{ventricle}(t)$. To define the elastance function for the ventricle, we follow the approach from Mynard et al., except we modify their formula so that our elastance function is exactly periodic \cite{Mynard12}. Specifically, the time dependent equation for the elastance of the ventricle on the interval $[0,T]$ is given by
\begin{align}
    E_{\text{ventricle}}(t) &= k\,\frac{g_1(t)}{1+g_1(t)} \left(\frac{1}{1+g_2(t)} - \frac{1}{1 + g_2(T)} \right)+E_{\text{min}}
\end{align}
where 
\begin{align}
\label{eq:gs}
g_i(t) = \left( \frac{t}{\tau_i}\right)^{m_i}, \quad i = 1,2.
\end{align}
The minimum and maximum elastances are $E_{\text{max}}= 1/C_{\text{systole}}$ and $E_{\text{min}} = 1/C_{\text{diastole}}$ and $k$ is a normalization factor defined as follows:
\begin{align}
    k = \frac{E_{\text{max}}-E_{\text{min}}}{\text{max}_{t \in [0,T]}\left[\frac{g_1(t)}{1+g_1(t)} \left(\frac{1}{1+g_2(t)} - \frac{1}{1 + g_2(T)} \right)\right]}.
\end{align}
 The parameters $\tau_1$ and $\tau_2$ appearing in equation \eqref{eq:gs} are the systolic and diastolic time constants, respectively. These values, along with $m_1$ and $m_2$, control the transition between the minimum and maximum elastance values for the single functioning ventricle.

The resistor elements in our models describe connections between compliance chambers. In this framework, compliance chambers $i$ and $j$ have two connections between them. The connection from $i$ to $j$ contains resistance $R_{ij}$ in series with a diode oriented from $i$ to $j$. The connection from $j$ to $i$ is set up analogously. If $R_{ij} = R_{ji}$, the compliance chambers are effectively connected by a single resistor and flow is allowed to move in both directions. To model a valve with no leak that allows flow from $i$ to $j$, we just set $R_{ij}$ equal to the (typically small) resistance of the valve, and $R_{ji}$ equal to infinity.  This setup, in which we separate the flows in the two possible directions, makes it possible also to model leaky valves (by making both resistances finite but unequal), and it has further benefit in the simulation of oxygen transport. In practice, we work with the reciprocal of the resistance, i.e.~the conductance, denoted $G_{ij} = 1/R_{ij}$. The flow (volume per unit time) from chamber $i$ to chamber $j$, denoted $Q_{\text{ij}}$, is assumed to obey Ohm's law,
\begin{align}
Q_{ij} = S(P_i,P_j)\,G_{ij}\,(P_i - P_j),
\label{eq:ohm}
\end{align}
where $S(P_i,P_j)$ is an indicator function that models the valve. This function equals one if $P_i > P_j$ and zero otherwise.

The equations of motion for the blood flow model follow from conservation of volume in each compliance chamber. Using our notation given above and defining $N$ to be the number of compliance chambers, we have
\begin{align}
\label{eq:conserve_volume}
    \frac{dV_i}{dt} = \sum_{j=1}^{N}(Q_{ji} - Q_{ij}), \quad i = 1, \ldots, N.
\end{align}
Substituting equations \eqref{eq:comp_chamber} and \eqref{eq:ohm} into \eqref{eq:conserve_volume}, we obtain:
\begin{align}
\label{eq:motion}
    \frac{d}{dt}(C_i P_i)= \sum_{j=1}^{N}(S(P_i,P_j)\,G_{ij} + S(P_j,P_i)\,G_{ji})(P_j - P_i), \quad i = 1, \dots, N.
\end{align}
This set of equations, one for each compliance chamber, comprise the blood flow model. Our model for Fontan circulation has four compliance chambers corresponding to the major vessel networks: the systemic arteries, systemic veins, pulmonary arteries, and pulmonary veins. The systemic organs, lungs, aortic valve, tricuspid valve, and Fontan connection are taken to be resistance elements. 

The fenestration between the systemic and pulmonary veins is taken to be a nonlinear resistor that depends on its cross-sectional area and the magnitude of flow through it. Let $Q_\text{Fe}$ denote the fenestration flow and $A_0$ denote its cross-sectional area. Positive fenestration flow corresponds to flow from the systemic venous chamber to the pulmonary venous chamber. An application of Bernoulli's equation on either side of the fenestration results in  
\begin{align}
\label{eq:fenres1}
   P_{\text{sv}}-P_{\text{pv}}=\frac{\rho}{2A_0}|Q_\text{Fe}|Q_\text{Fe},
\end{align}
where $P_{\text{sv}}$ and $P_{\text{pv}}$ are the systemic venous and pulmonary venous pressures respectively. Equation \eqref{eq:fenres1} reveals the resistance of the fenestration, as a function of the magnitude of the fenestration flow and the size: 
\begin{align}
    R_{\text{Fe}} = \frac{\rho}{2A_0}|Q_\text{Fe}|.
\end{align}
We add an additional term to this resistance, denoted $R_\text{visc}$:
\begin{align}
    R_{\text{Fe}} = R_\text{visc}+\frac{\rho}{2A_0}|Q_\text{Fe}|.
\end{align}
 This additional term corresponds to a small viscous resistance and is practically important since it allows for the fenestration conductance to remain finite when $Q_\text{Fe} = 0$. The formula for fenestration conductance is 
\begin{align}
   G_{\text{Fe}} =  \frac{1}{R_\text{visc}+\frac{\rho}{2A_0}|Q_{\text{Fe}}|},
   \label{eq: fen_conductance}
\end{align}
which is used in \eqref{eq:motion} for the conductance between the systemic and pulmonary venous compliance chambers when the fenestration is open. Note that the equations in \eqref{eq:motion} are nonlinear because of  the functions $S(P_i,P_j)$. The fenestration conductance $G_{\text{Fe}}$ introduces an additional nonlinearity which needs be carefully handled in the numerical method, to be described below.

\subsection{Oxygen Model}
\label{subsec:o2model}
In our model for oxygen transport, each compliance chamber has a time-dependent oxygen concentration denoted $[O_2]_i$. This variable corresponds to a volumetric concentration so that $V_i[O_2]_i$ is the volume of oxygen\footnote{The \textit{volume of oxygen} is the volume that the amount of oxygen
 in question would occupy as a gas under standard conditions, i.e.,
 atmospheric pressure and zero degrees Celsius.  For example, one
 mole of oxygen (or any ideal gas) has a volume of 22.4 liters
 under standard conditions.  Note that volumetric oxygen concentration
 is dimensionless (volume/volume), and an amazing coincidence is that
 fully saturated blood has about the same volumetric oxygen concentration
 as the atmosphere of Earth, namely 1/5.} in the $i$th compliance chamber. We also define sources and sinks of oxygen along the resistance elements between compliance chambers, denoted by the set of parameters $M_{ij}$. These parameters are nonzero only for the connections between the arteries and veins on both sides of the circulation. Conservation of oxygen in the circulation model is described by the following set of equations:
\begin{align}
    \frac{d}{dt}(V_i[\text{O}_2]_i) = \sum_{\substack{j=1 \\ j\neq i}}^N (Q_{ji}[\text{O}_2]_j - Q_{ij}[\text{O}_2]_i + M_{ji}) \quad i = 1, \ldots, N.
    \label{eq:oxygen_diff_eq}
\end{align}
The first term in the sum on the right hand side is the rate of oxygen entering chamber $i$ from $j$, using the concentration from chamber $j$. The second term is the rate of oxygen leaving chamber $i$, and the third term, if nonzero, is either a source $(M_{ji} > 0)$ or sink ($M_{ji} < 0$). For the connection between the systemic arteries and systemic veins, the baseline value of oxygen consumption, denoted $M_\text{sa,sv}$, is determined from Fick's law, assuming the concentration in the systemic veins is $60\%$ of the concentration in the systemic arteries \cite{Hijazi92}:
\begin{align}
M_{\text{sa,sv}} = -0.4 \, Q_\text{s} \, \gamma.    
\label{eq:ficks}
\end{align}
In computing $M_\text{sa,sv}$ from equation \eqref{eq:ficks}, $Q_\text{s}$ is taken to be the cardiac output from our model Fontan circulation without the fenestration, and $\gamma = 0.22$ corresponds to the volumetric oxygen concentration in blood when it is fully saturated \cite{Barnea94}. The source of oxygen from the lungs, denoted $M_\text{pa,pv}$, is chosen in a time-dependent fashion so that the blood flowing into the pulmonary veins from the pulmonary arteries is fully saturated. This requirement implies the following equation for $M_\text{pa,pv}$: 
\begin{align}
M_{\text{pa,pv}} = Q_\text{p}(\gamma - [\text{O}_2]_{\text{pa}}),
\end{align}
where $Q_\text{p}$ is the pulmonary flow and $[O_2]_\text{pa}$ is the oxygen concentration in the pulmonary arteries. This source term ensures that blood flowing into the pulmonary veins from the pulmonary arteries has a volumetric oxygen concentration $\gamma = 0.22$. In the presence of a fenestration, however, this stream of fully oxygenated blood mixes with the systemic venous blood that arrives in the pulmonary venous compartment via the fenestration, with the result that blood in the pulmonary venous compartment is not fully saturated.

\subsection{Numerical methods}
\label{subsec:num_methods}

In this section, we describe the numerical approximation schemes for the blood flow and oxygen transport models. Let $n$ denote the time step index and $\delta t > 0$ denote the time step. For the blood flow equations, we use the backward Euler scheme:
\begin{align}
    \frac{C_{i}^nP_{i}^n - C_{i}^{n-1}P_{i}^{n-1}}{\delta t} & = \displaystyle\sum_{j=1}^{N} (S(P_i^n,P_j^n)G_{ij}^n + S(P_j^n,P_i^n)G_{ji}^n)(P_{j}^n - P_{i}^n).
    \label{eq:num_blood_flow}
\end{align}
There are two sources of nonlinearity in these equations: the valve state function $S$, which depends on the pressures, and the fenestration conductance, which depends on the flow. Since the valve state function $S$ can only take on the values 0 or 1, our approach for solving the nonlinear system \eqref{eq:num_blood_flow} is to guess the value for each $S$ variable, to solve the resulting linear system for the pressures, and finally to check whether the pressure values obtained are consistent with the assumed values of $S$. If there is any inconsistency, we change the state of those $S$ values that were inconsistent with the pressures and try again.  This is repeated until the pressures and all of the valve states have stopped changing, i.e., until the nonlinear system \eqref{eq:num_blood_flow} has been solved. Our initial guess for this process is taken to be values at the previous time, which for almost all time steps is the correct choice. To deal with the nonlinearity in the fenestration conductance, we perform a fixed-point iteration on the fenestration flow, with the flow at the previous time step used as the initial guess. This iteration helps to avoid non-physical oscillations in the calculated fenestration flow waveform. Our numerical discretization for the oxygen transport equations is the following:
\begin{equation} 
\label{eq:num_o2} 
\frac{[O_{2}]_{i}^nV_{i}^n - [O_{2}]_{i}^{n-1}V_{i}^{n-1}}{\delta t} = \displaystyle\sum_{j = 1 \atop j\ne i}^N ([O_{2}]_{j}^{n-1} Q_{ji}^n -[O_{2}]_{i}^{n-1}Q_{ij}^n + M_{ji}).
\end{equation}
The scheme is forward Euler in terms of the concentrations. We use the flow values calculated at the next time step since we have them available via Ohm's law after updating the pressures with equation \eqref{eq:num_blood_flow}.

\section{Results and discussion}

In this section, we present and discuss results from our numerical simulations. In all cases, we use 100 time steps per cardiac cycle, and simulations are run for 4000 cardiac cycles. As mentioned in Subsection \ref{subsec:num_methods}, 10 fixed point iterations per time step are performed on equation \eqref{eq:num_blood_flow} to avoid nonphysical oscillations caused by the nonlinear fenestration conductance. The duration of the cardiac cycle is set to $T = 0.016$ minutes and is determined by stroke volume and cardiac output data from Fontan patients reported in \cite{Liang14} (see Table \ref{tab:calibrated_variables}). The duration of these simulations is chosen to guarantee that all hemodynamic and oxygen transport variables reach periodic steady states in all cases. Although the hemodynamic variables typically reach a periodic steady state within 10-20 cycles, the oxygen transport variables take many more cycles to reach a periodic steady state. The blood volume used in all simulations is 5 liters, and initial oxygen concentrations in all compliance chambers are set to $\gamma = 0.22$. The choice of initial oxygen concentrations has no effect on the periodic steady state that is eventually reached. Mean values of variables reported below correspond to an average taken over the last three cardiac cycles of the simulation.

\begin{table}
\begin{center}
\label{TestTable}
\begin{tabular}{l*{6}{c}r}
\hline
Parameters            & Resistance ($R$) & Dead Volume ($V_\text{d}$) & Compliance ($C$) \\
\hline
  Units   & mmHg min L$^{-1}$ & L & L mmHg$^{-1}$   \\
\hline 
S           & 20.78 & - & -   \\
P           & 0.5517 & - & -   \\
Ao           & 0.01 & - & -   \\
Tr           & 0.01 & - & -   \\
Fo           & 0.01 & - & -   \\
SA          & - & 0.7051 & $7.333\times 10^{-4}$   \\
PA         & - & 0.0930 & 0.00412   \\
SV        & - & 2.869 & 0.0990  \\
PV        & - & 0.1475 & 0.01 \\

\hline
\end{tabular}
\caption{Parameters for the circulation model. Abbreviations: S, systemic organs; P, pulmonary; Ao, aortic valve; Tr, tricuspid valve; Fo, Fontan connection; SA, systemic arteries; PA, pulmonary arteries; SV, systemic veins; PV, pulmonary veins. }
\label{table:parms1}
\end{center}
\end{table}

\subsection{Model calibration with a closed fenestration}

We first describe the calibration of our model using hemodynamic data from Fontan patients, with the goal of deriving a baseline set of parameters. The model is calibrated, with the fenestration closed, to a data set provided by Liang et al.  \cite{Liang14}. Tables \ref{table:parms1} and \ref{table:parms2} display the parameters used within the calibrated model. Table \ref{tab:calibrated_variables} shows clinical variables calculated from our calibrated model compared to those reported by Liang et al. \cite{Liang14}.  Calibration was done by manual tuning with normal circulation parameters used as initial guesses. Variables that incorporate blood volume are indexed to a body surface area of 1.5 m$^2$.  In general, variables from our model match well with the clinical data.

\begin{figure}[h!]
\begin{center}
  \includegraphics[scale=0.55,trim=0 0 -20 0]{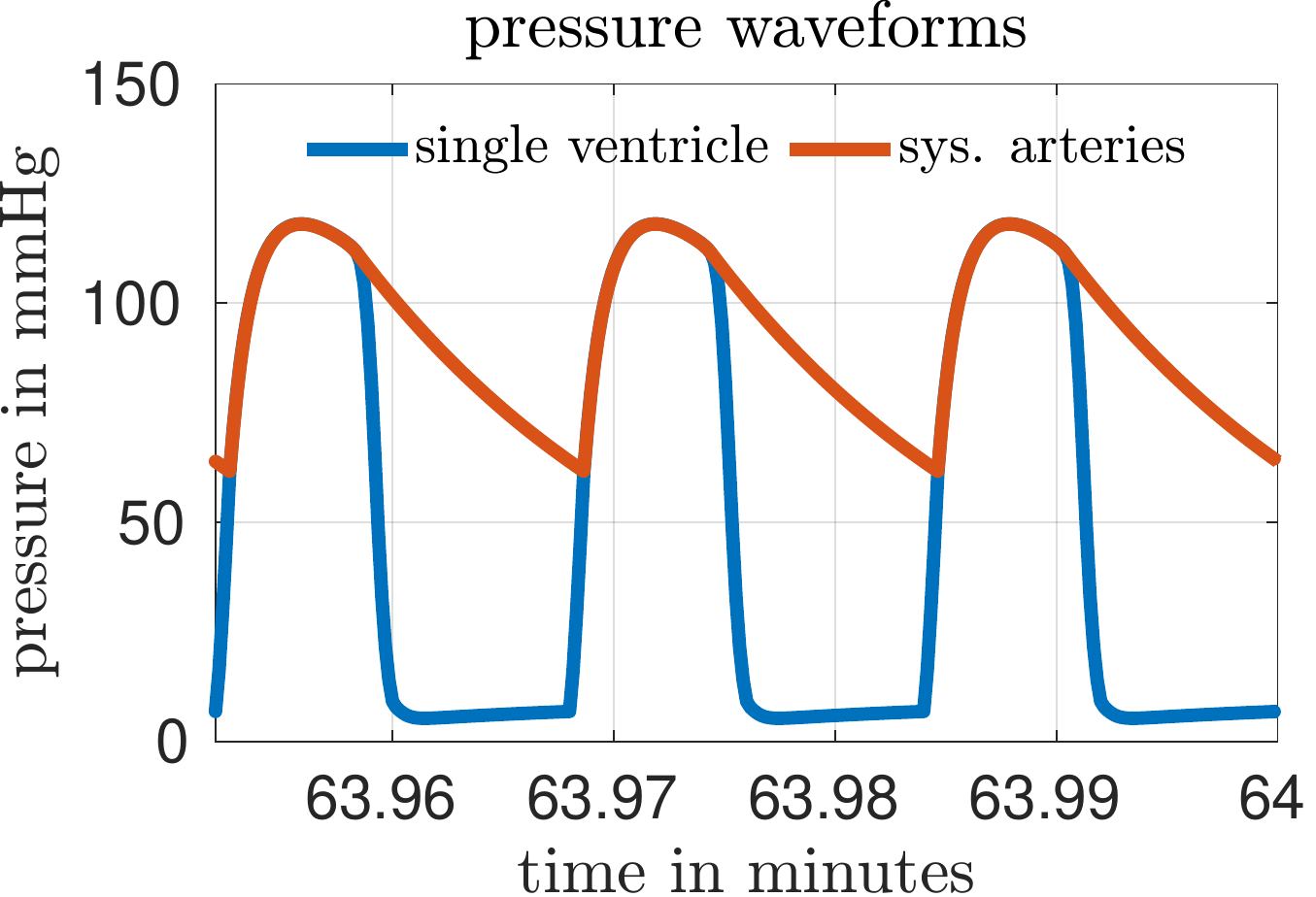}
  \includegraphics[scale=0.55]{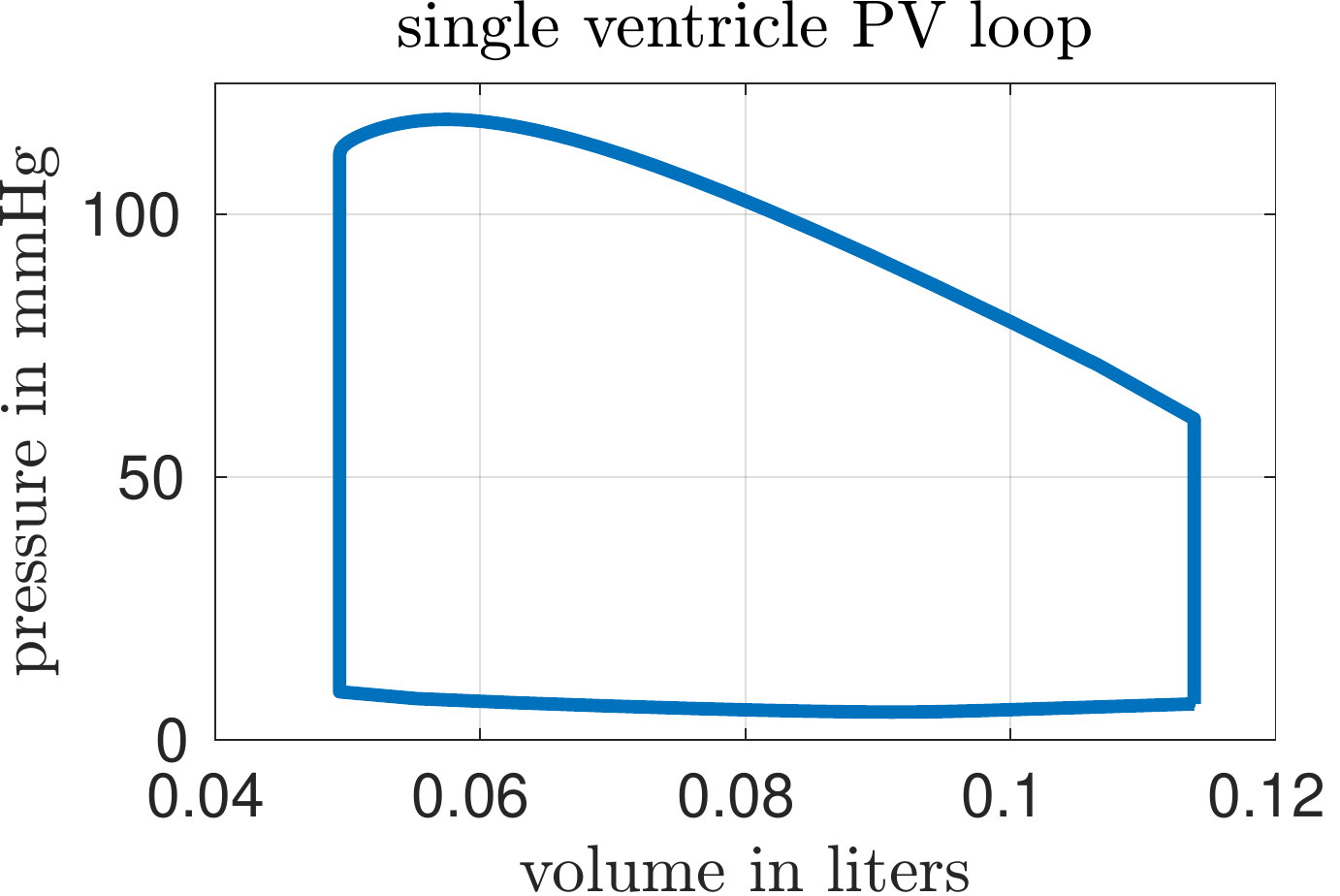}
\caption{Results from our calibrated model with a closed fenestration. Three pressure waveforms from the end of the simulation are shown in the left panel and a pressure-volume loop for the single ventricle is shown in the right panel.}
\label{fig:press_and_pv}
\end{center}
\end{figure}

Based on a report by Ohuchi, the calibrated variables from our model correspond to a ``late-surviving'' and ``excellent-surviving'' Fontan patient \cite{Ohuchi17}. ``Late-surviving'' refers to a follow-up of at least 15 years after the Fontan operation, and ``excellent-surviving'' is characterized by a central venous pressure of 10 mmHg, a cardiac index of 2.6 L min$^{-1}$ m$^{-2}$, an end diastolic volume index of 70 mL m$^{-2}$, and a stroke volume index of 55 mL m$^{-2}$ (refer to Table 1 in \cite{Ohuchi17}). These hemodynamic values align well with those from our calibrated model. Figure \ref{fig:press_and_pv} shows results from our calibrated model with a closed fenestration. The left panel depicts pressure waveforms for the ventricle and systemic arteries over three cardiac cycles. The right figure shows the ventricular pressure-volume loop. 

\begin{table}
\begin{center}
\begin{tabular}{l*{6}{c}r}
\hline
Parameters              & Symbol         & Units   &  Ventricle \\
\hline
Minimal elastance       & $E_\text{min}$ & mmHg L$^{-1}$  & 79.52         \\
Maximal elastance       & $E_\text{max}$ & mmHg L$^{-1}$ & 5232          \\
systolic exponent    & $m_1$          & -       & 1.32            \\
diastolic exponent     & $m_2$          & -       & 27.4            \\
Systolic time constant  & $\tau_1$       & minutes & 0.269$\times T$         \\
Diastolic time constant & $\tau_2$       & minutes & 0.452$\times T$         \\
Dead volume             & $V_\text{d}$   & L       & 0.028         \\
Period of heartbeat     & $T$            & minutes & 0.016  \\
\hline
\end{tabular}
\caption{Parameters for the time varying ventricular compliance in the heart model.}
\label{table:parms2}
\end{center}
\end{table}

\begin{table}
\begin{center}
\begin{tabular}{l*{6}{c}r}
\hline
variable                                                        & our model & clinical data reported in \cite{Liang14} \\
\hline
cardiac index (L\,min$^{-1}$\,m$^{-2}$) & 2.686  & 2.9, 2.1      \\
stroke volume index (mL m$^{-2}$)                  & 42.98  & 39,40         \\
end diastolic volume index (mL m$^{-2}$)        & 75.90  & 72, 76        \\
end systolic volume index (mL m$^{-2}$)         & 32.93  & 33, 36        \\
end systolic pressure (mmHg)                                 & 118.1  &  124$^\star$            \\
end diastolic pressure (mmHg)                                & 6.827  & 6.6           \\
vena cava mean pressure (mmHg)                                  & 9.349  & 8             \\
pulse pressure (mmHg)                                           & 56.39  & 54             \\
systemic artery systolic pressure (mmHg)                        & 118.1  & 124           \\
systemic artery diastolic pressure (mmHg)                       & 61.66  & 70            \\
systemic artery mean pressure (mmHg)                            & 93.07  & 88            \\
pulmonary artery mean pressure (mmHg)                                 & 9.308 & 9       \\
\hline
\end{tabular}
\caption{Hemodynamic variables calculated from our calibrated model with a closed fenestration, compared to the clinical data reported by Liang et al. \cite{Liang14}. Based on a report by Ohuchi, our model's vena cava mean pressure, cardiac index, end diastolic volume index, and stroke volume index are all within the range of a ``late-surviving'' and ``excellent-surviving'' Fontan patient \cite{Ohuchi17}. The indexing in the first four variables assumes a body surface area of $1.5 $ m$^2$. $^\star$Clinical data was not available for the ventricular end systolic pressure, so we compare our model results with the reported systemic arterial systolic pressure.}
\label{tab:calibrated_variables}
\end{center}
\end{table}

\subsection{Blood and oxygen transport with an open fenestration}

In this section, we explore the effect of an open fenestration on both hemodyamic and oxygen transport variables. As described in Subsection \ref{subsec:o2model}, the baseline value for oxygen consumption in our models is calculated from Fick's law via equation \eqref{eq:ficks}, assuming 60\% saturation in the systemic veins and using the systemic flow $Q_\text{s}$ from the calibrated model with the closed fenestration. These choices result in a baseline oxygen consumption value $M_{\text{sa,sv}} = -0.354552$ L min$^{-1}$. 

\begin{figure}[h!]
\begin{center}
  \includegraphics[scale=0.55,trim=0 0 -20 0]{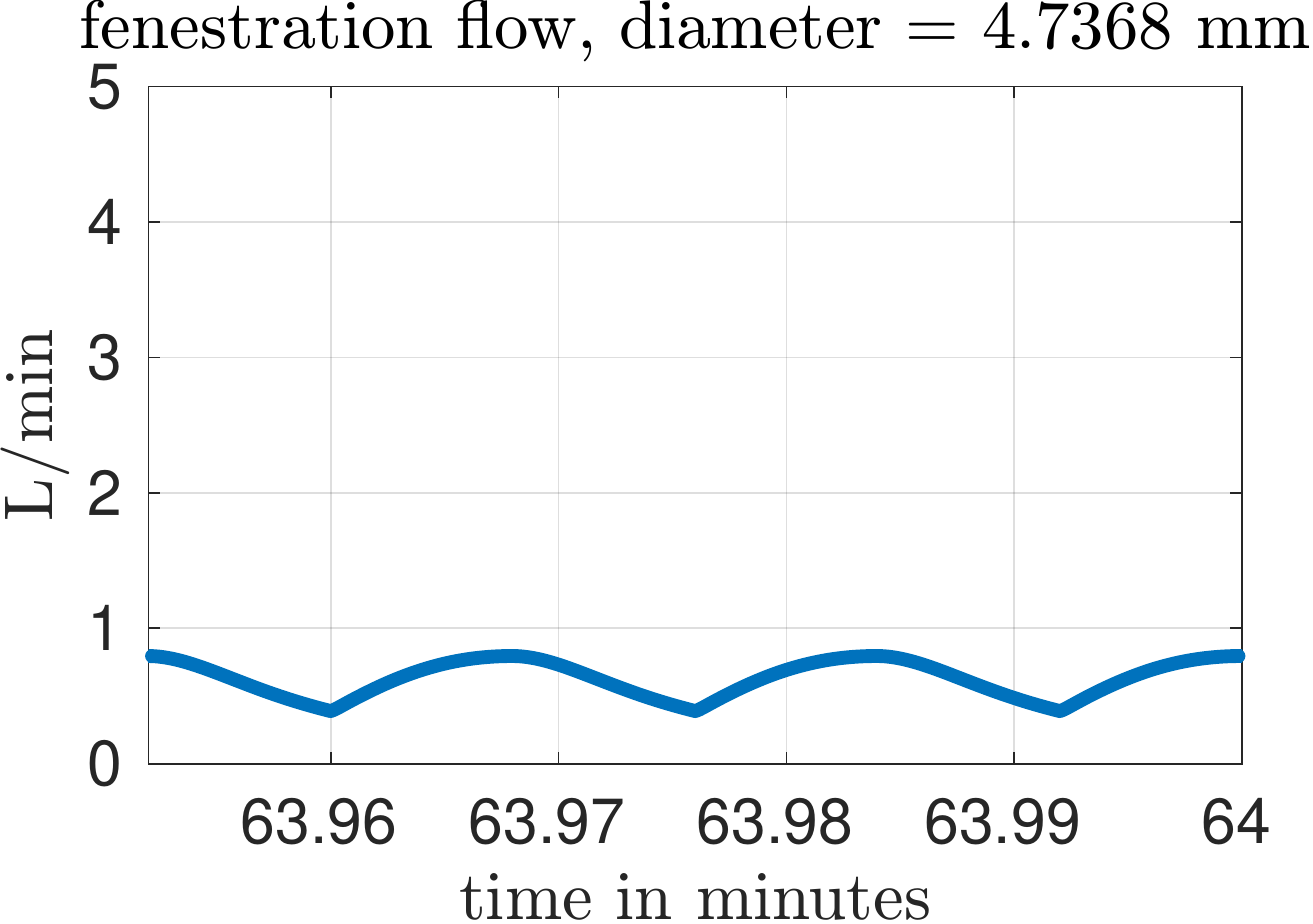}
  \includegraphics[scale=0.55]{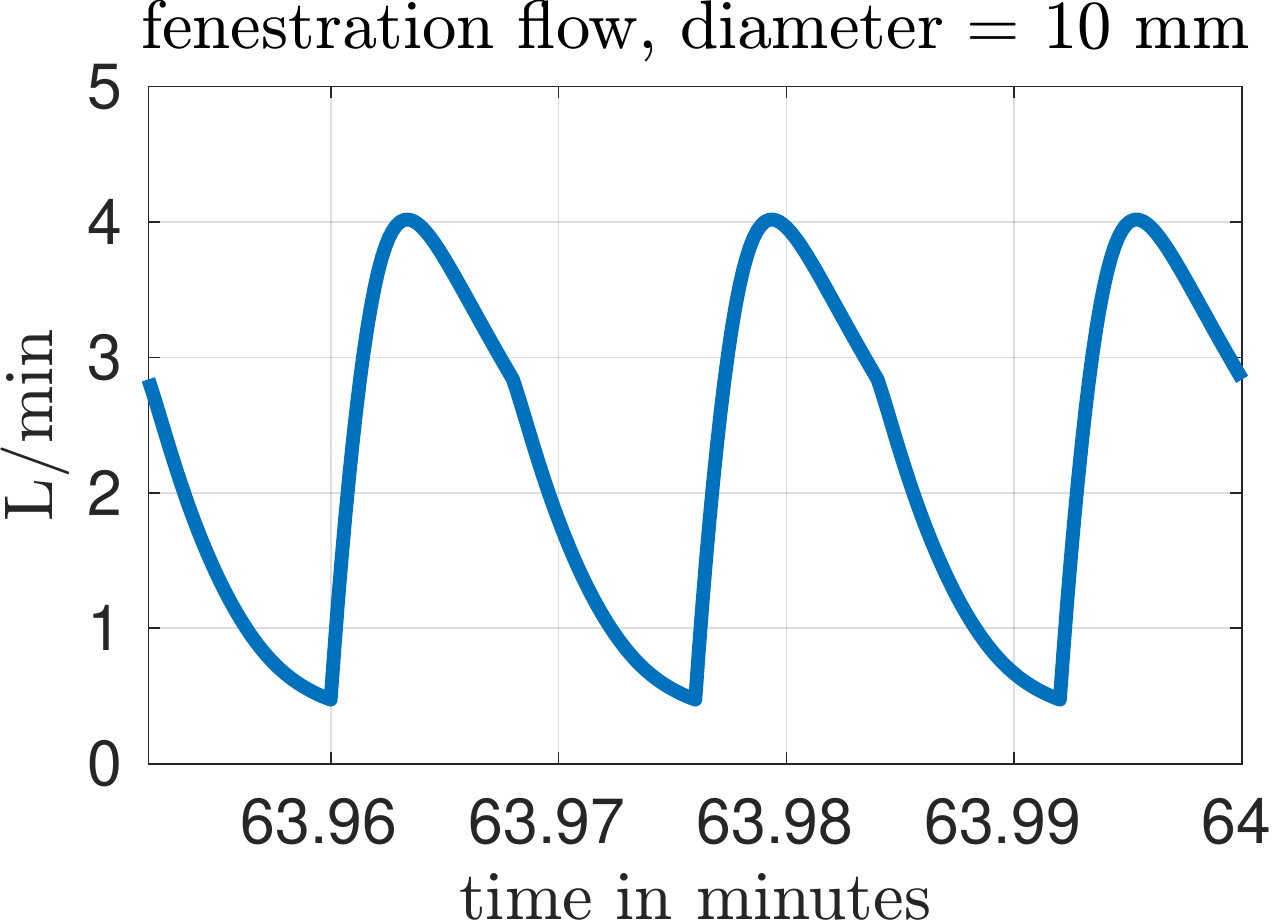}
  \caption{Fenestration flow waveforms for two different fenestration sizes and our baseline pulmonary vascular resistance value $R_\text{P} =$ 0.5517 mmHg min L$^{-1}$. The waveform in the left panel corresponds to a fenestration with diameter equal to 4.7368 mm, and the waveform in the right panel corresponds to a diameter of 10 mm.}
  \label{fig:results0}
\end{center}  
\end{figure}

Figure \ref{fig:results0} shows fenestration flow waveforms for two different fenestration sizes. The left panel corresponds to a fenestration with diameter equal to 4.7368 mm and the right panel corresponds to a diameter of 10 mm. The pulmonary vascular resistance is set to our baseline value of $R_\text{P} = 0.5517$ mmHg min L$^{-1}$. Although the model allows for bi-directional flows, the fenestration flow throughout the cardiac cycle is exclusively from the systemic veins to the pulmonary veins, with more flow occurring in the larger fenestration. Note that the fenestration flow is roughly proportional to the cross-sectional area of the fenestration, since it is approximately multiplied by four when the diameter of the fenestration approximately doubles.  This makes sense, since flow proportional to area means that velocity is constant, as we would expect from the Bernoulli relation for a given pressure difference.  At larger fenestration sizes, however, the pressure difference would be expected to diminish, and the flow would eventually saturate at some level that is independent of the fenestration size.

\begin{figure}[h!]
\centering
  \begin{subfigure}[b]{0.47\textwidth}
  \centering
  \includegraphics[scale=0.55]{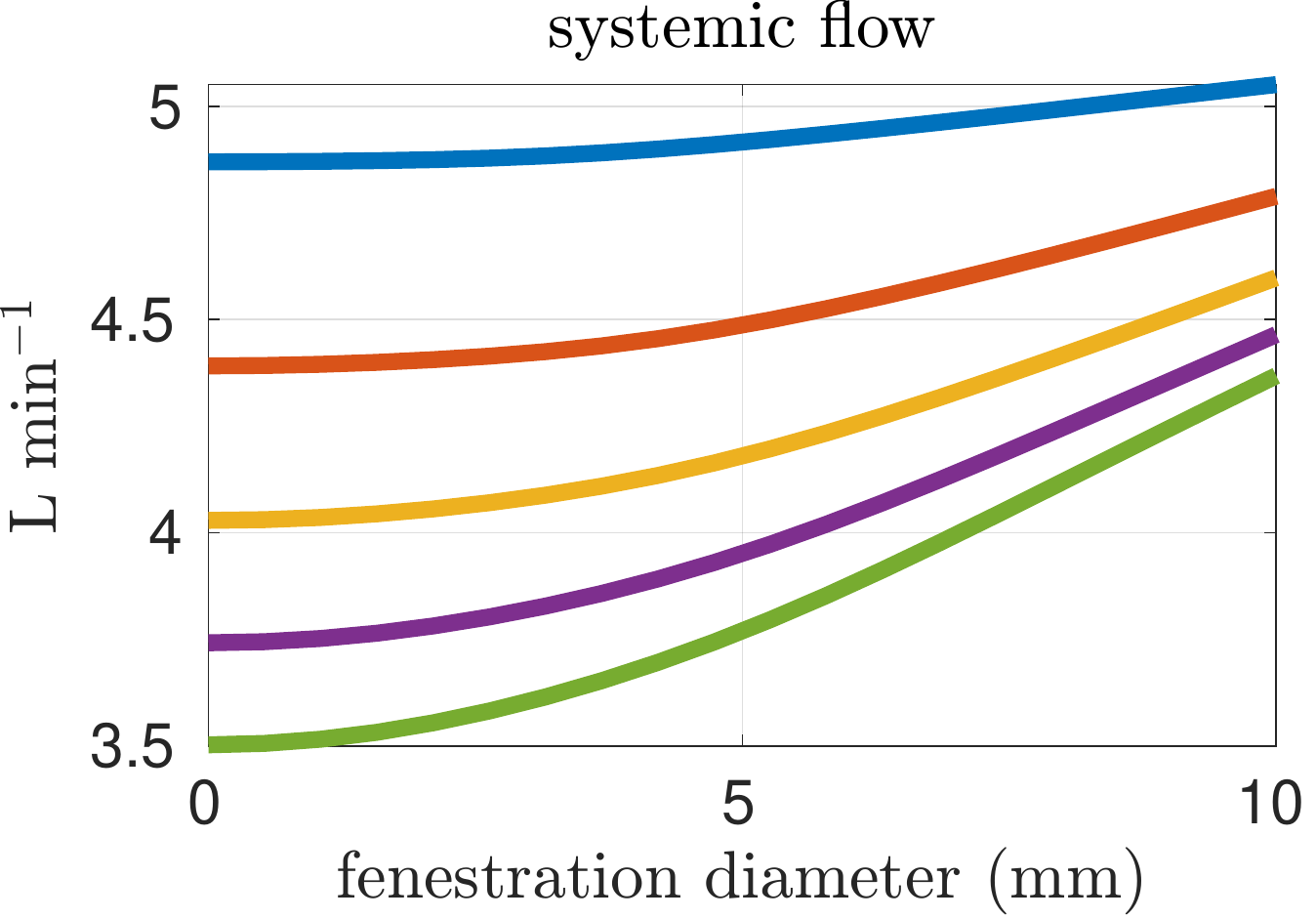}
  \caption{}
  \end{subfigure}
   \begin{subfigure}[b]{0.47\textwidth}
  \centering
  \includegraphics[scale=0.55]{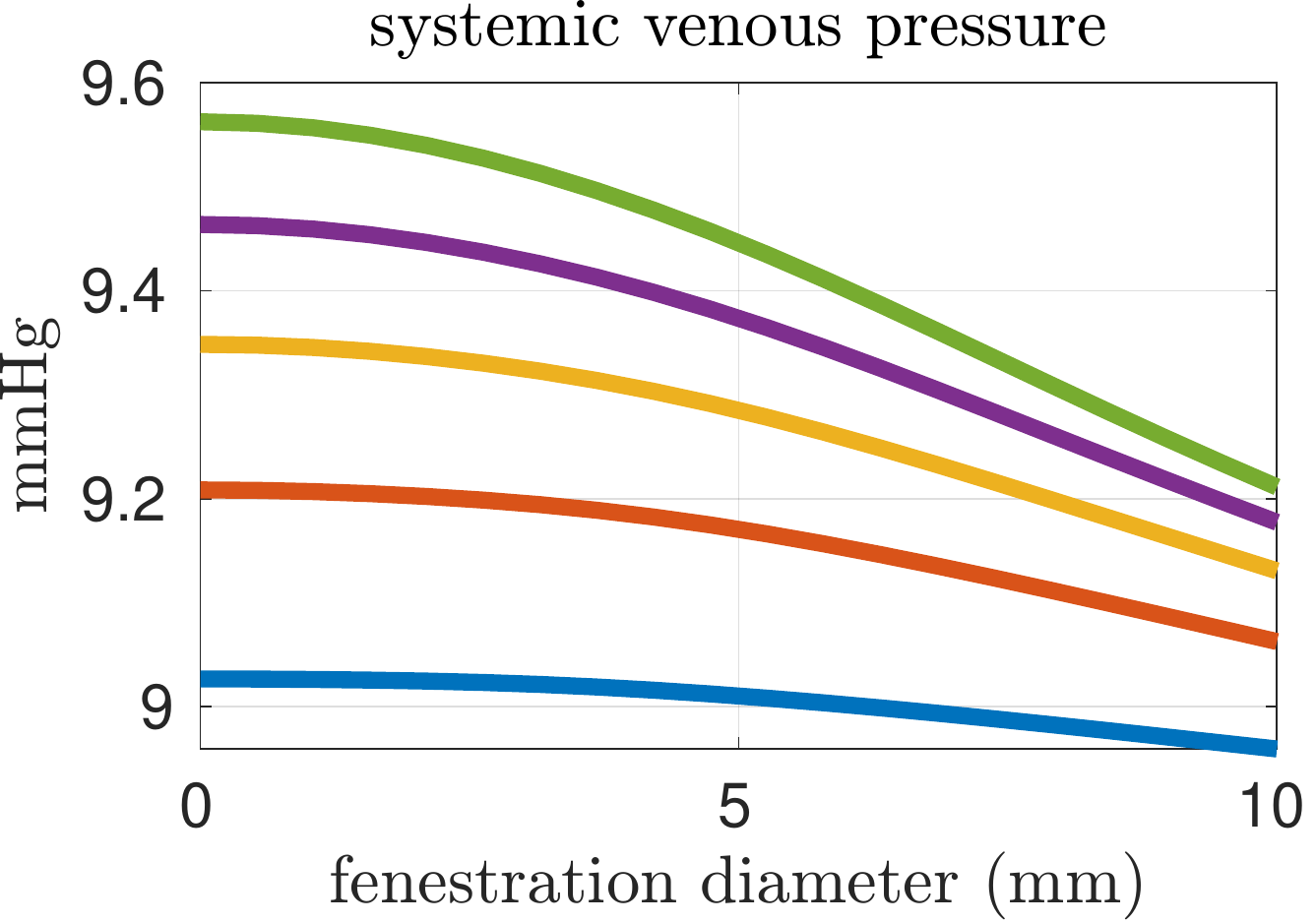}
  \caption{}
  \end{subfigure}\\
  \vspace{0.3cm}
  \begin{subfigure}[b]{0.47\textwidth}
  \centering
  \includegraphics[scale=0.55]{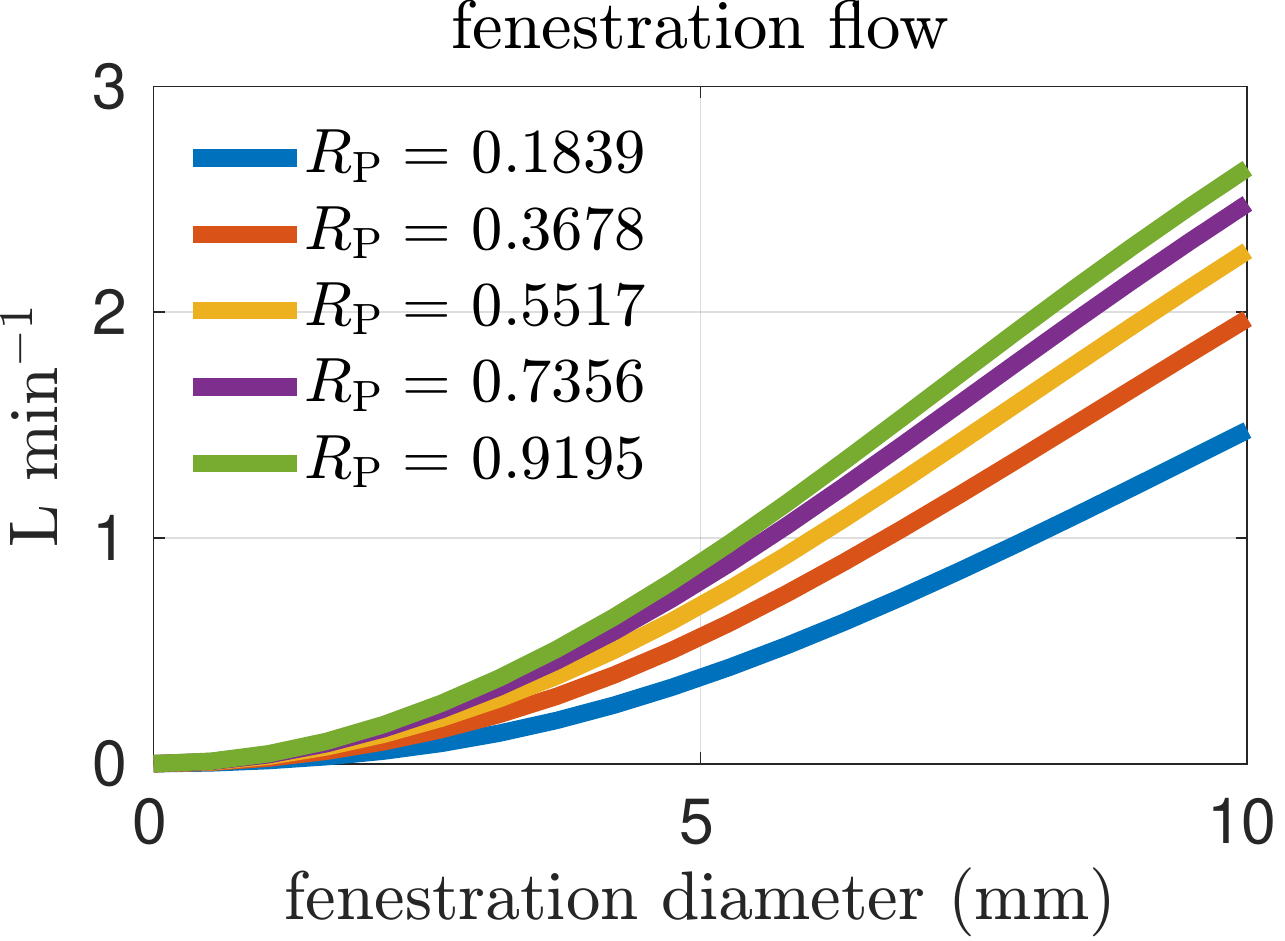} 
  \caption{}
  \end{subfigure}
  \begin{subfigure}[b]{0.47\textwidth}
  \centering
  \includegraphics[scale=0.55]{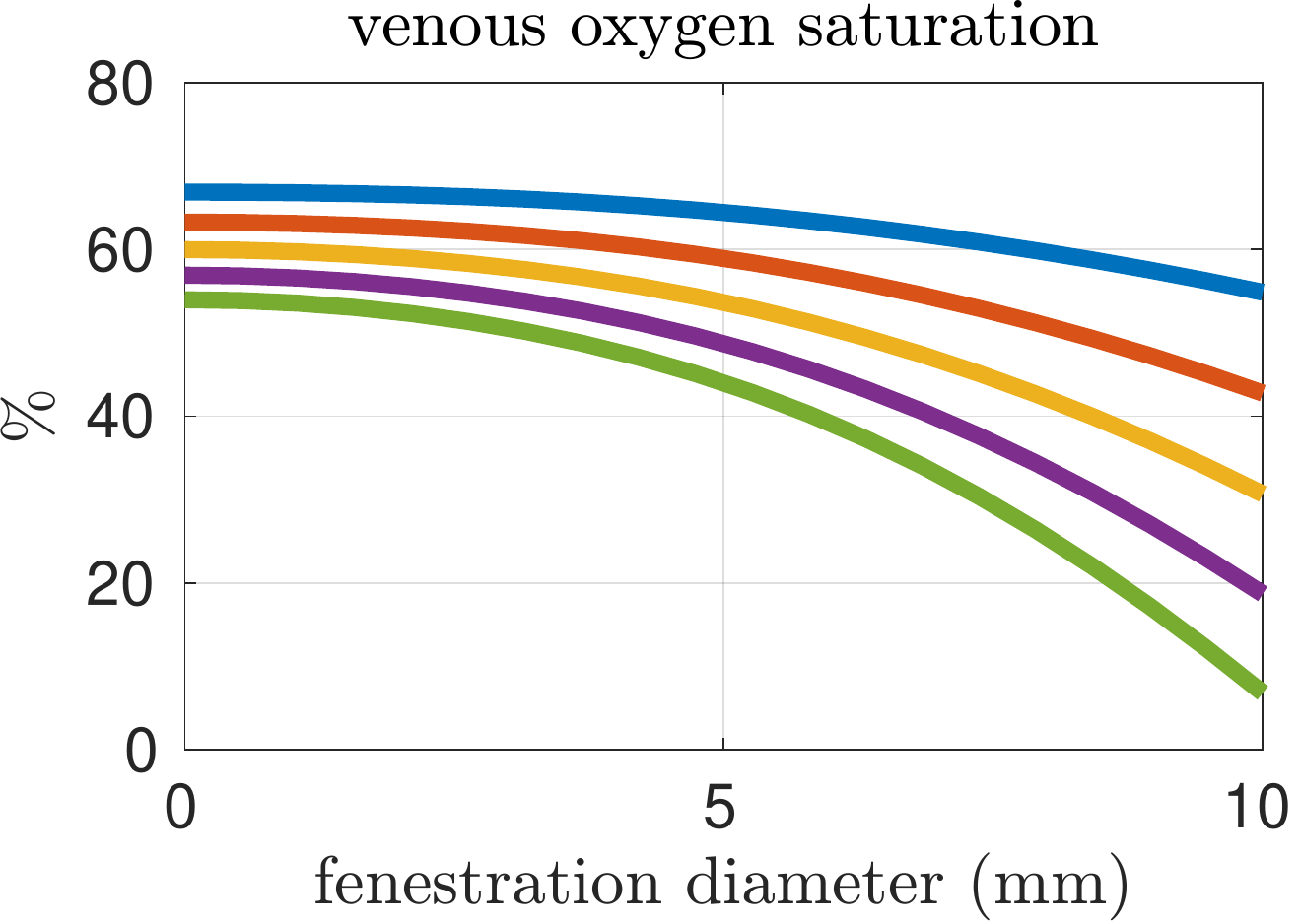}
  \caption{}
  \end{subfigure}\\
  \vspace{0.3cm}
  \begin{subfigure}[b]{0.47\textwidth}
  \centering
  \includegraphics[scale=0.55]{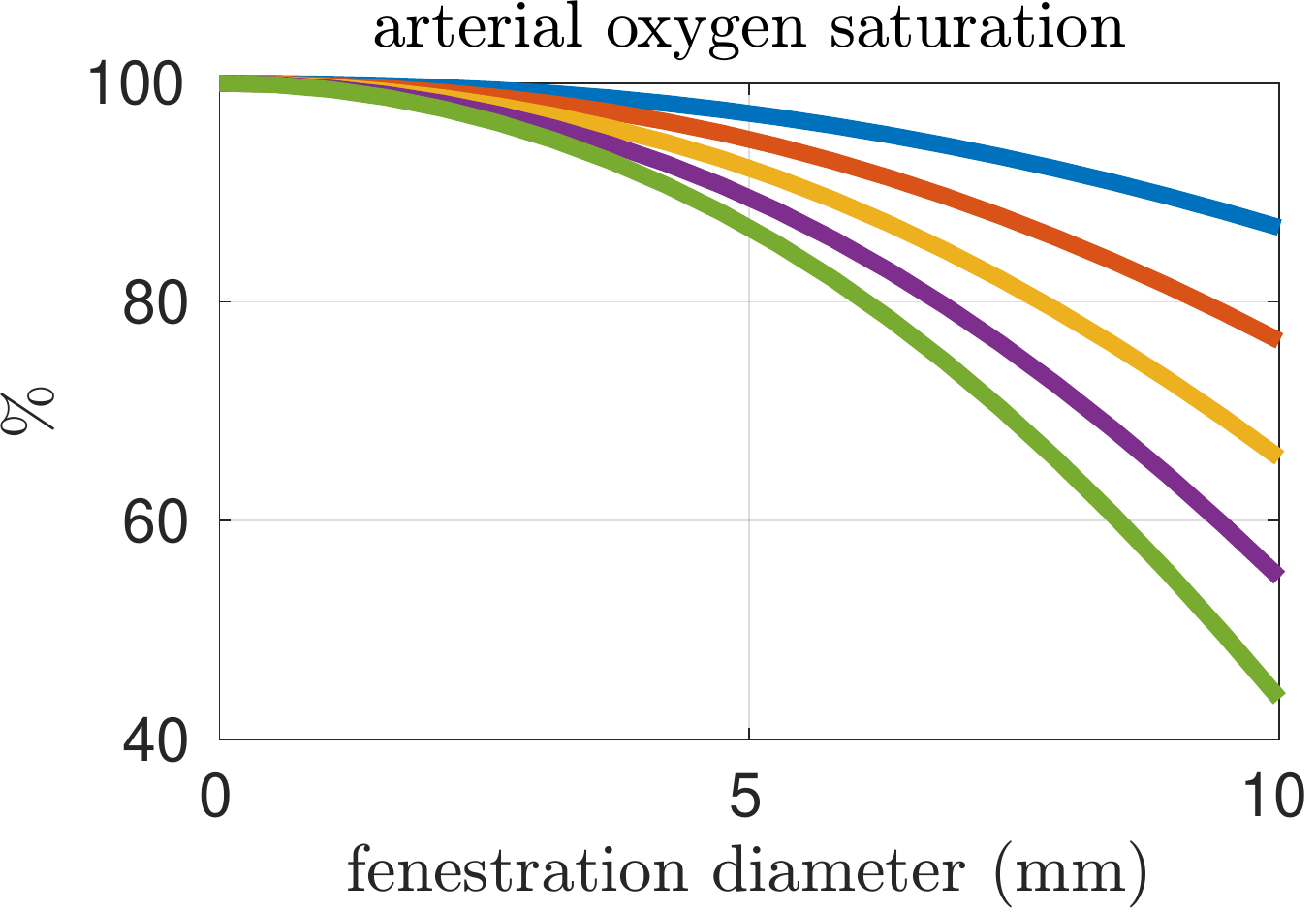}
  \caption{}
  \end{subfigure}
  \begin{subfigure}[b]{0.47\textwidth}
  \centering
  \includegraphics[scale=0.55]{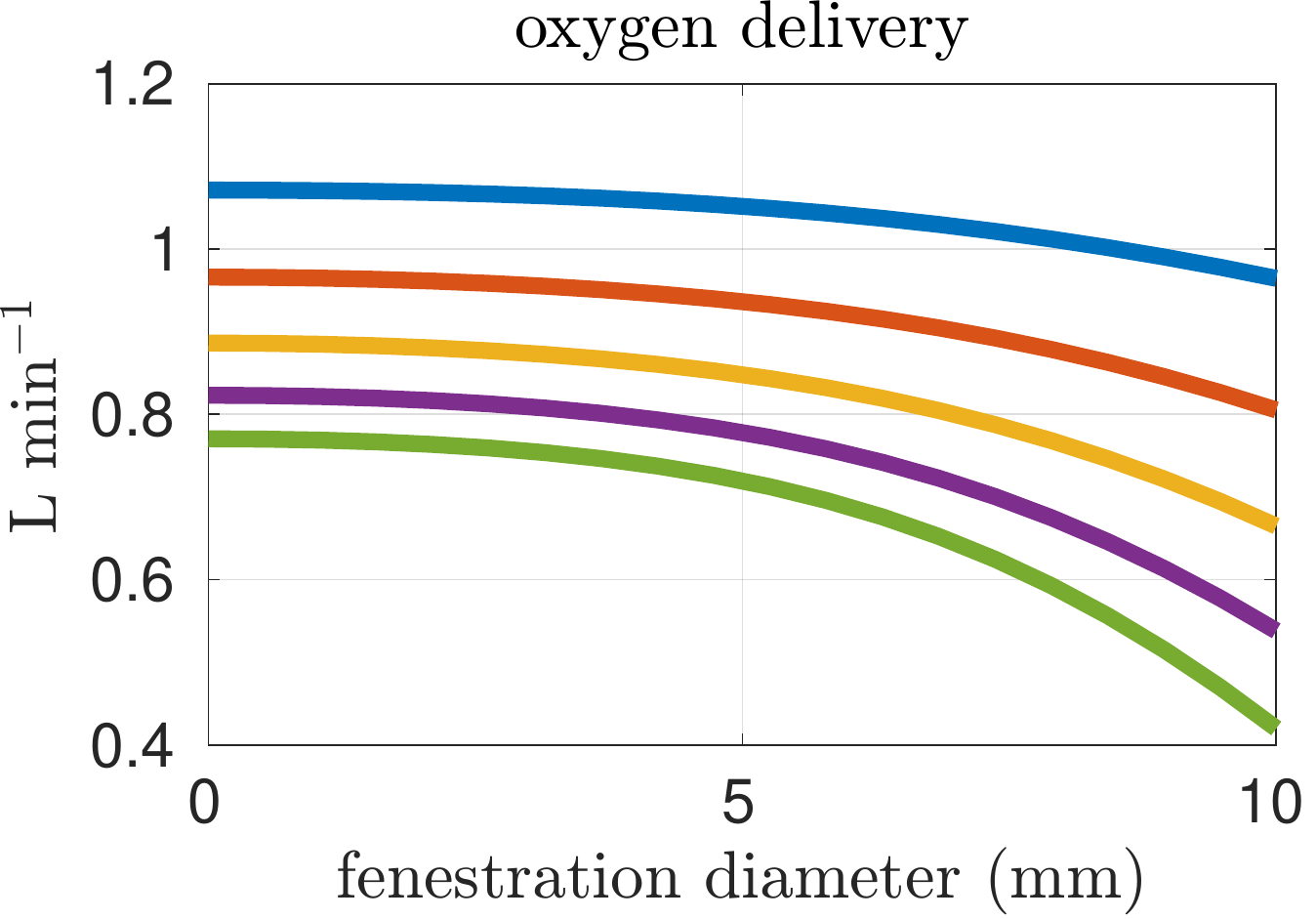}
  \caption{}
  \end{subfigure}
\caption{Results corresponding to oxygen consumption equal to -0.354552 L min$^{-1}$. }
\label{fig:results1}
\end{figure}

Figure \ref{fig:results1} shows results from our model with the oxygen consumption parameter $M_{\text{sa,sv}}$ set to the baseline value. To explore the effect of pulmonary resistance on  hemodynamic and oxygen transport variables, we consider five values for the resistance that are equally spaced around the baseline value of $R_\text{P} = 0.5517$ mmHg min L$^{-1}$. All variables are plotted as functions of the fenestration diameter. The range of fenestration diameters considered here, from 0-10 mm, contains the range of diameters seen in clinical reports, see e.g.~\cite{Grosse13}. The upper two panels show systemic flow and systemic venous pressure. As expected, an increase in fenestration size leads to an increase in systemic flow and a decrease in systemic venous pressure. Changes in these variables are more pronounced for higher pulmonary vascular resistances. The middle two panels show the mean flow through the fenestration and the systemic venous oxygen saturation. In general, fenestration flow is larger for higher pulmonary vascular resistances, and it increases monotonically with fenestration diameter. The values for mean fenestration flow from our models are physiologically reasonable when compared to flows measured in patients via phase contrast cardiac magnetic resonance imaging \cite{Grosse13}.  Blood flowing through the fenestration is not reoxygenated, so larger fenestration sizes and pulmonary vascular resistances result in lower systemic venous saturations. Note that the systemic venous oxygen saturation for the closed fenestration and for the case $R_\text{P} = 0.5517$ mmHg min L$^{-1}$ is 60\%, which is precisely how the baseline oxygen consumption parameter for this set of experiments was determined. Another reason we examine systemic venous saturation is to ensure it is positive for a given set of parameters, since there is nothing in the model to preclude this variable from going negative. These results confirm that venous saturations are positive for all parameter choices, although they are quite small for larger pulmonary resistances and fenestration sizes.  The lower two panels show systemic arterial oxygen saturation and oxygen delivery, the latter of which is defined as the product of systemic flow and arterial oxygen saturation. The trends in systemic arterial oxygen saturation are the same as for the systemic venous saturation. For oxygen delivery, all curves monotonically decrease with increasing fenestration diameter. A more dramatic decrease can be seen for larger pulmonary resistances. The monotonic nature of the oxygen delivery curve changes when we consider parameter values corresponding to a ``high-risk'' Fontan patient, to be described below.

\begin{figure}[h!]
\centering
  \begin{subfigure}[b]{0.47\textwidth}
  \centering
  \includegraphics[scale=0.55]{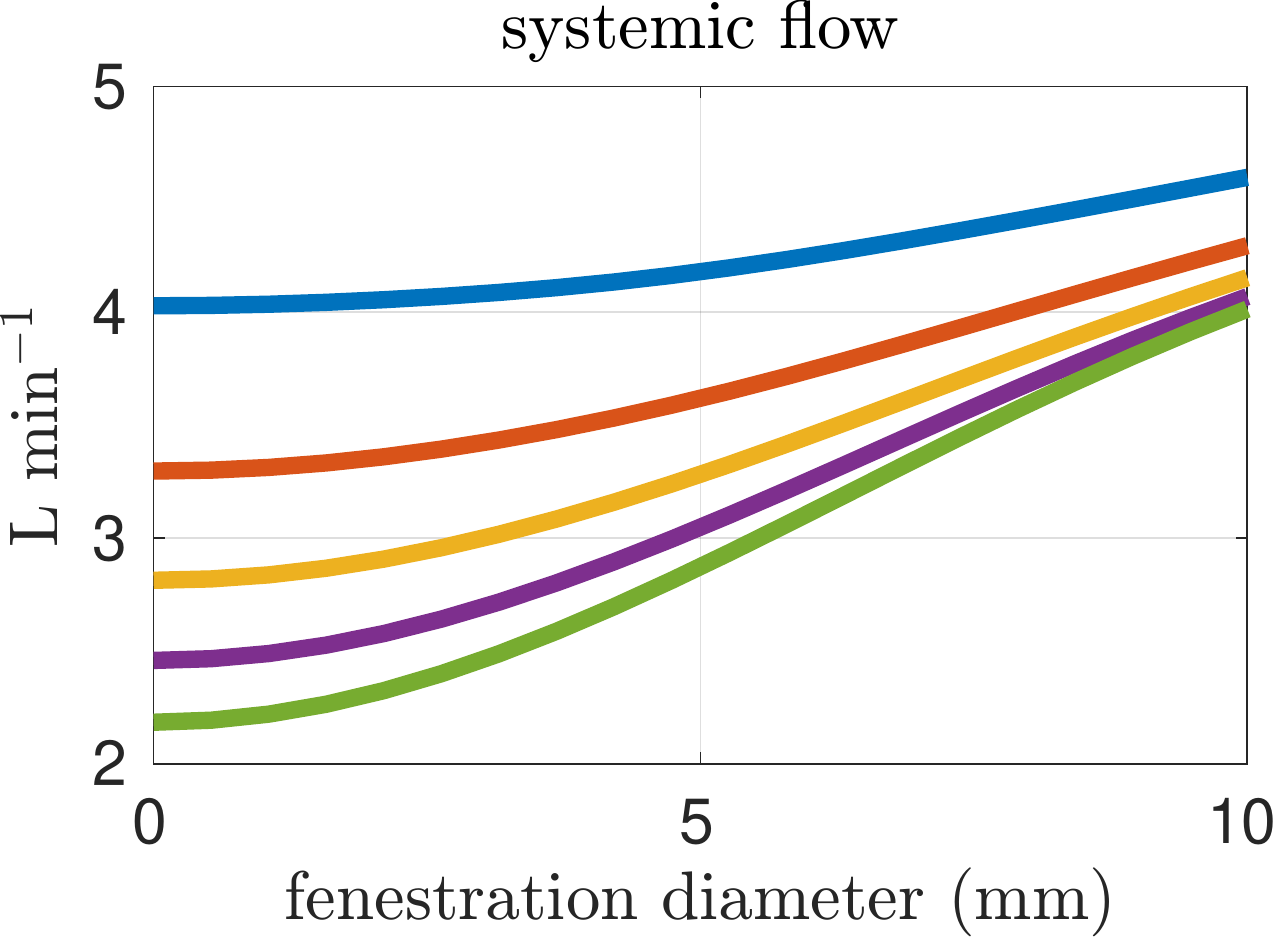}
  \caption{}
  \end{subfigure}
  \begin{subfigure}[b]{0.47\textwidth}
  \centering
  \includegraphics[scale=0.55]{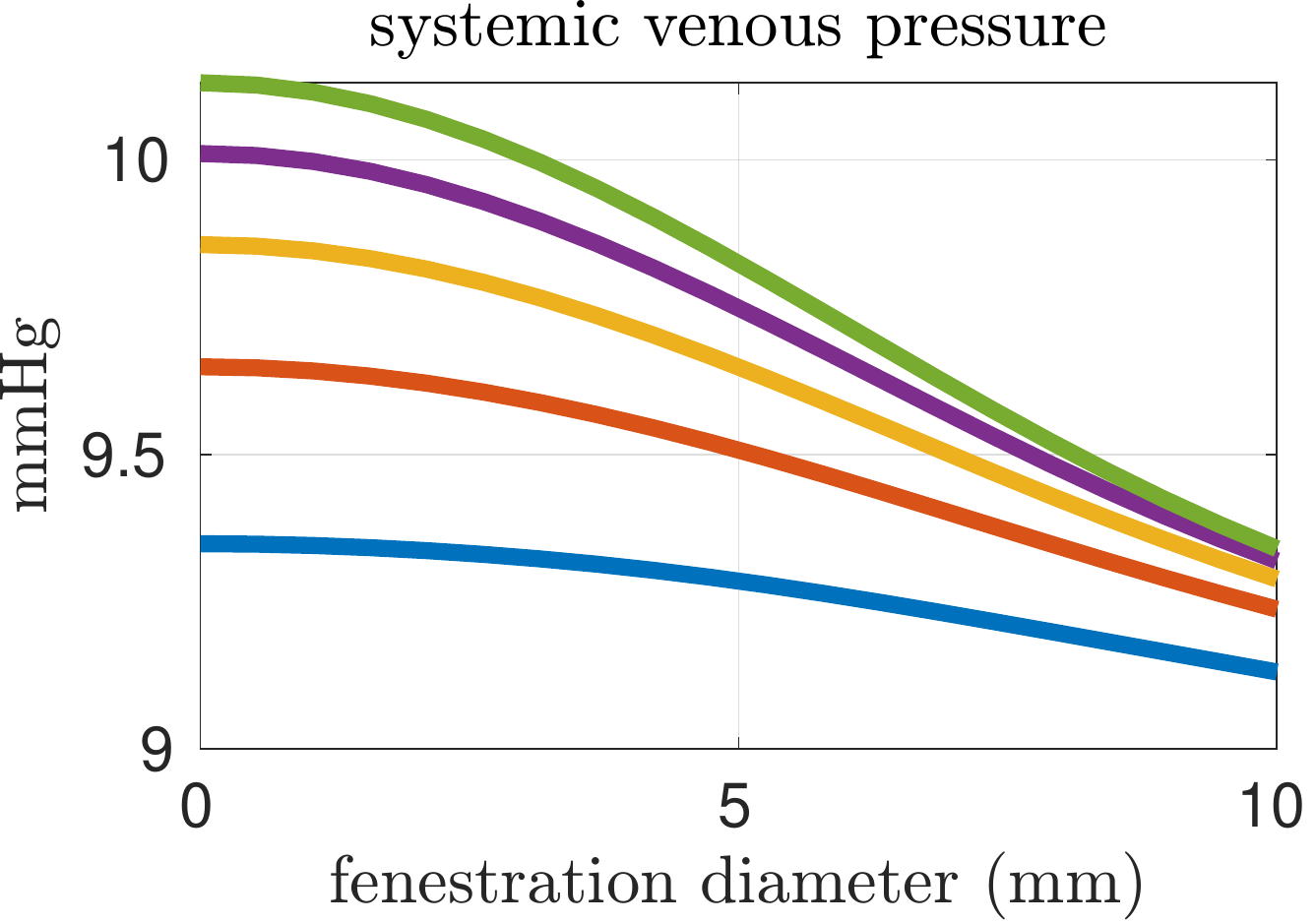}
  \caption{}
  \end{subfigure}\\
  \vspace{0.3cm}
  \begin{subfigure}[b]{0.47\textwidth}
  \centering
  \includegraphics[scale=0.55]{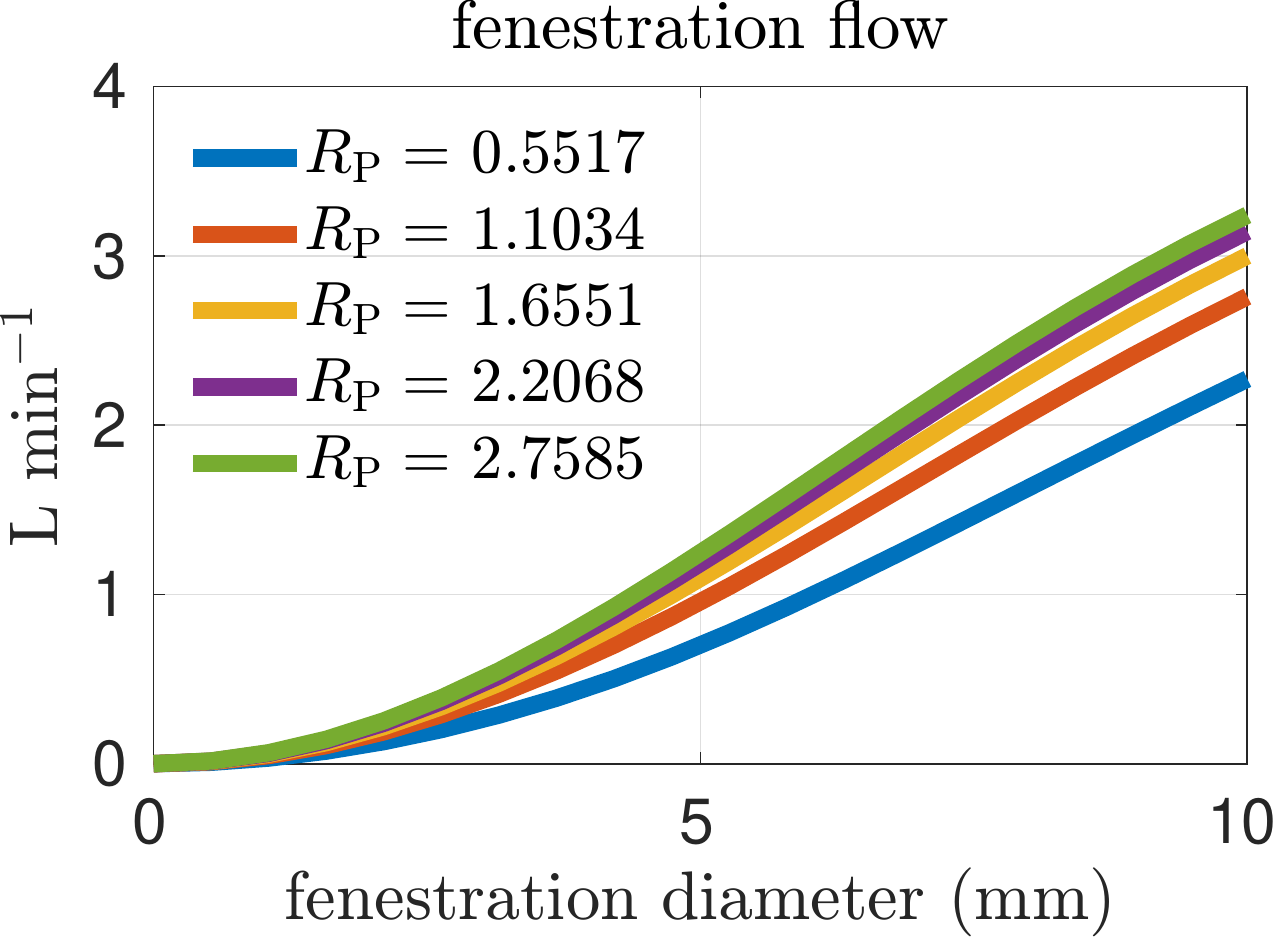}
  \caption{}
  \end{subfigure}
  \begin{subfigure}[b]{0.47\textwidth}
  \centering
  \includegraphics[scale=0.55]{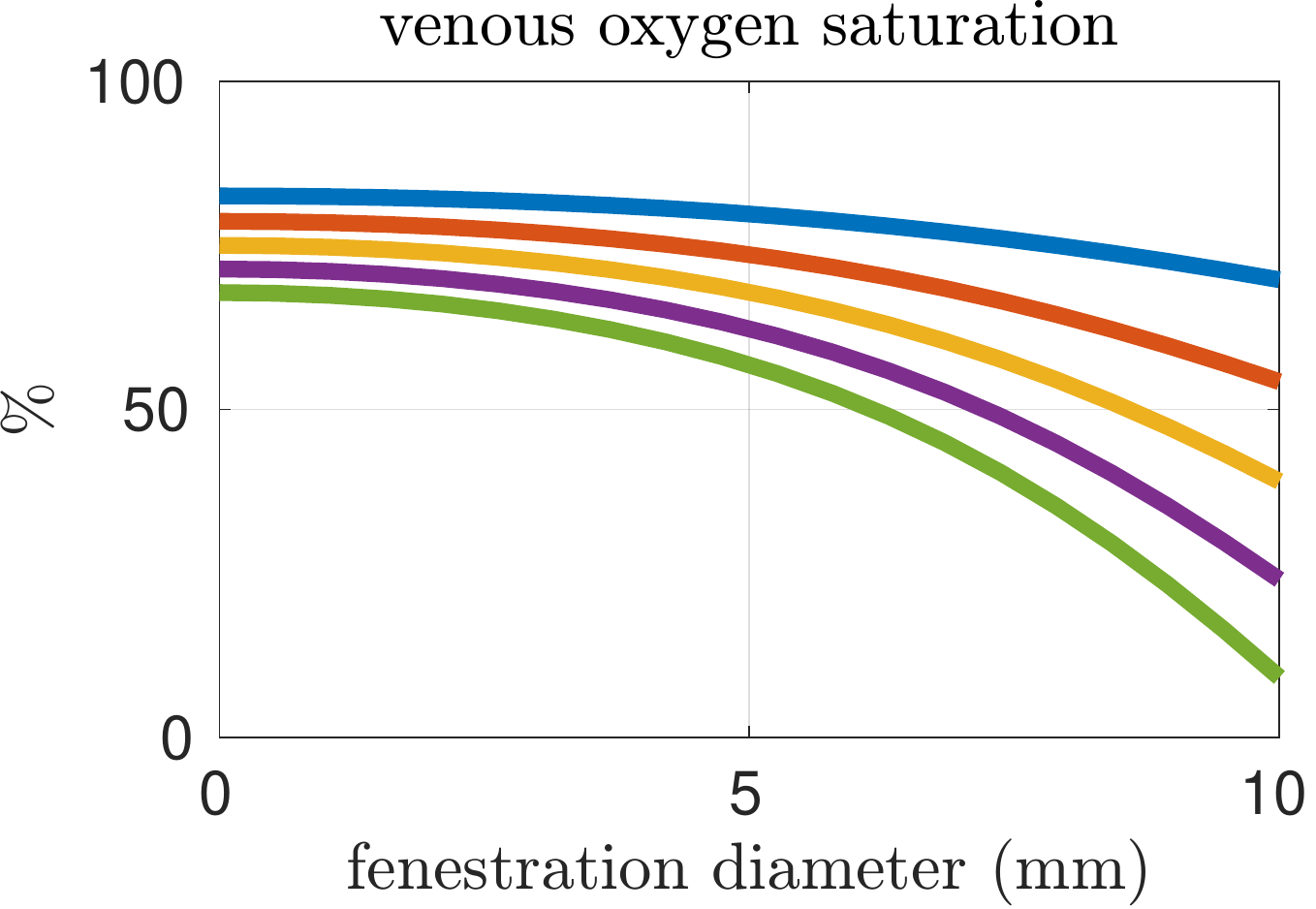}
  \caption{}
  \end{subfigure}\\
  \vspace{0.3cm}
  \begin{subfigure}[b]{0.47\textwidth}
  \centering
  \includegraphics[scale=0.55]{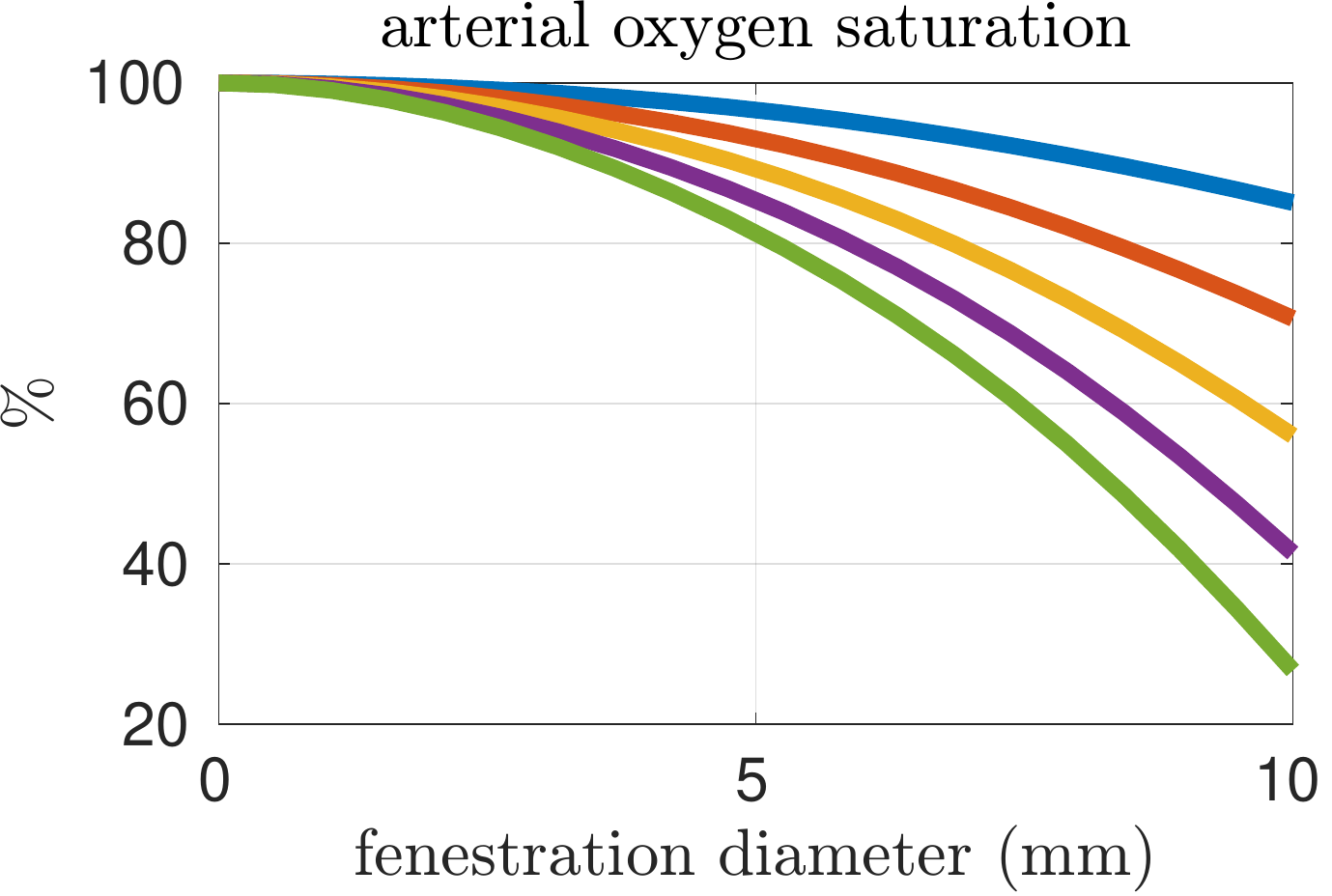}
  \caption{}
  \end{subfigure}
  \begin{subfigure}[b]{0.47\textwidth}
  \centering
  \includegraphics[scale=0.55]{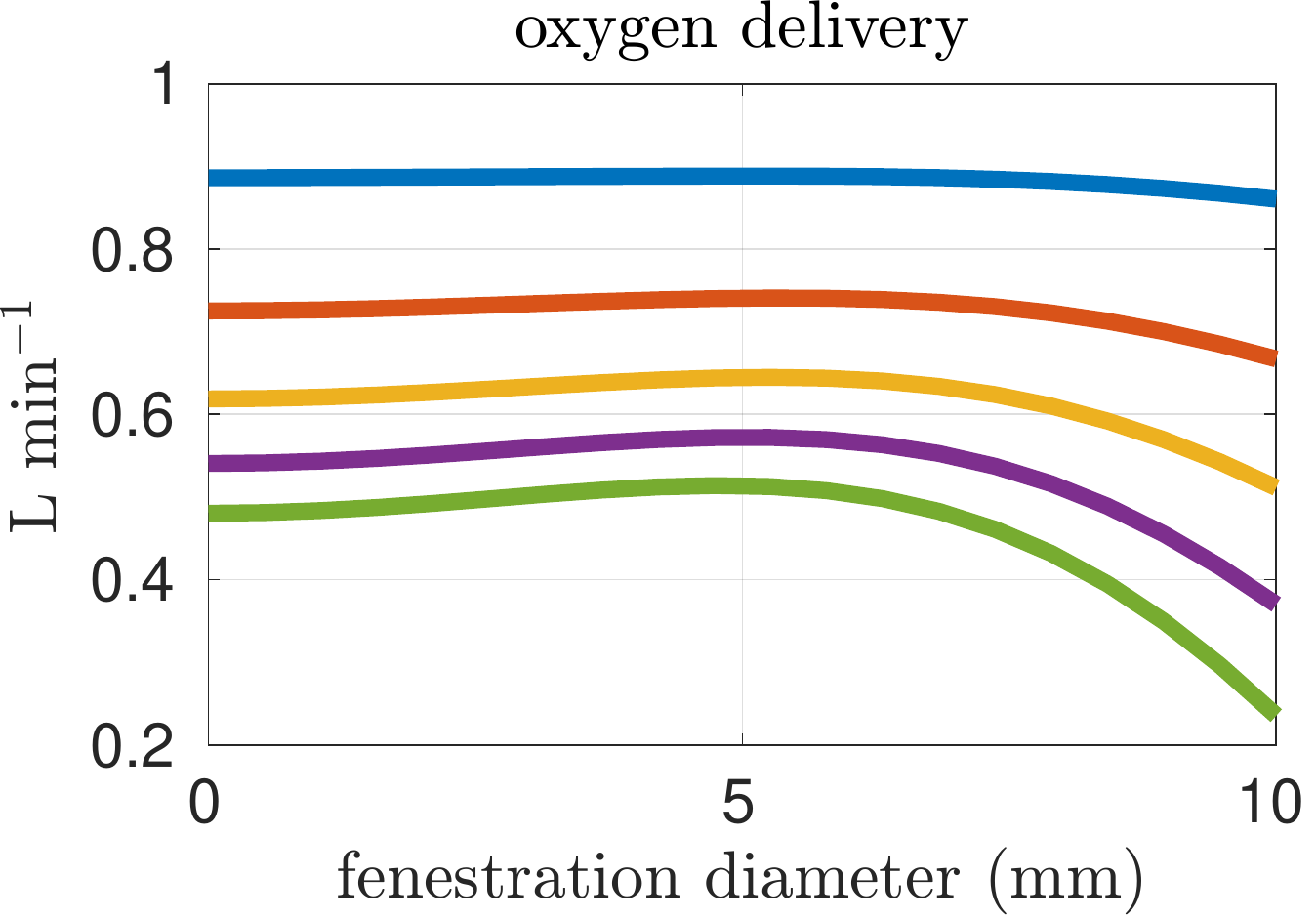}
  \caption{}
  \end{subfigure}
\caption{Results corresponding to oxygen consumption equal to -0.154552 L min$^{-1}$.}
\label{fig:results2}
\end{figure}

High pulmonary vascular resistance and low cardiac output have been identified as important risk factors for Fontan patients \cite{Lemler02, Egbe17}. More specifically, a pulmonary vascular resistance index\footnote{PVRI does not directly correspond to the $R_\text{P}$ parameter in our models. It is defined as the mean pulmonary artery pressure divided by the cardiac index.} (PVRI)  greater than 2 mmHg min L$^{-1}$ m$^{2}$ and a cardiac index less than 2.5 L min$^{-1}$ m$^{-2}$ is a criteria that has been used to characterize high-risk patients \cite{Egbe17}. Referring to Table \ref{tab:calibrated_variables}, the PVRI and cardiac index for our baseline model are 3.465 mmHg min L$^{-1}$ m$^{2}$ and 2.686 L min$^{-1}$ m$^{-2}$ respectively. According to Egbe et al., our baseline model aligns with the cohort containing more favorable Fontan physiology. In the next set of experiments, we consider pulmonary vascular resistance values that are larger than our baseline value in order to generate models that correspond to ``high-risk'' Fontan physiology. Figure \ref{fig:results2} shows analogous results to Figure \ref{fig:results1} except for pulmonary resistance values larger than our baseline value. Note that an increase in pulmonary vascular resistance substantially decreases the cardiac output. Thus, in these cases it is necessary to decrease the oxygen consumption (in these cases, we use $M_{\text{sa,sv}} = -0.154552$ L min$^{-1}$) so that venous saturations remain positive. 

In general, the fenestration has a more substantial impact on hemodynamics and oxygen transport for larger pulmonary vascular resistances. The trends seen here are the same as in Figure \ref{fig:results1}, except for oxygen delivery. In these high-risk cases (with larger pulmonary vascular resistance, smaller cardiac index, and smaller oxygen consumption), the delivery curve as a function of fenestration diameter is non-monotonic. This implies a range of fenestration sizes with larger oxygen delivery than the closed fenestration case, presumably because the increase in cardiac output, as a result of the open fenestration, is able to compensate for the decrease in systemic arterial oxygen saturation. In particular, there is an optimal fenestration diameter which maximizes oxygen delivery. The peaks in the oxygen delivery curves are more pronounced for the higher pulmonary resistances. In all cases, delivery either increases or is relatively constant as fenestration size initially increases. These results point towards the possibility of fenestrating high-risk patients to achieve lower pulmonary artery/systemic venous pressures and higher cardiac output without the cost of a decrease in oxygen delivery. 

\begin{figure}[h!]
\begin{center}
  \includegraphics[scale=0.55, trim=0 0 -20 0]{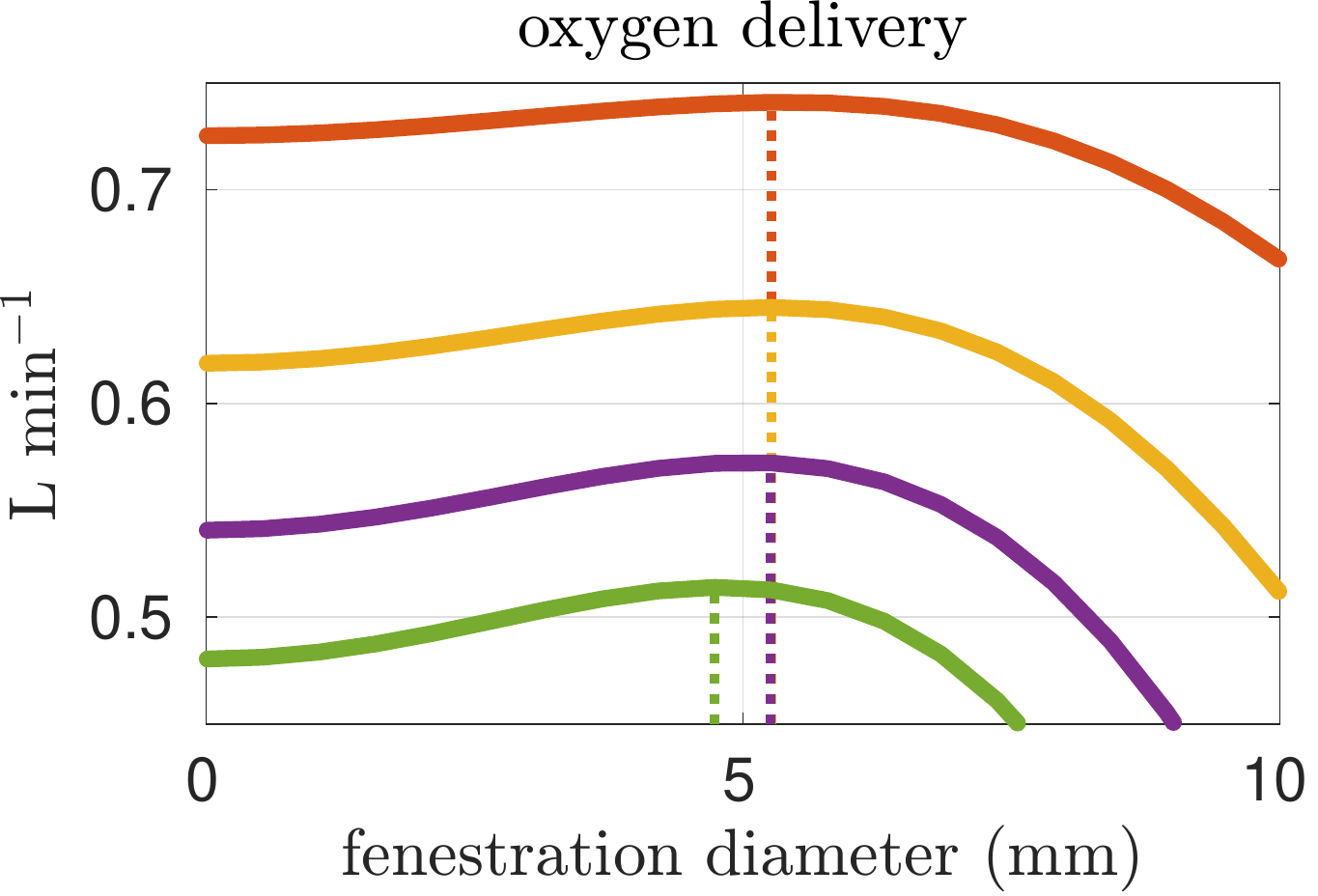}
  \includegraphics[scale=0.55]{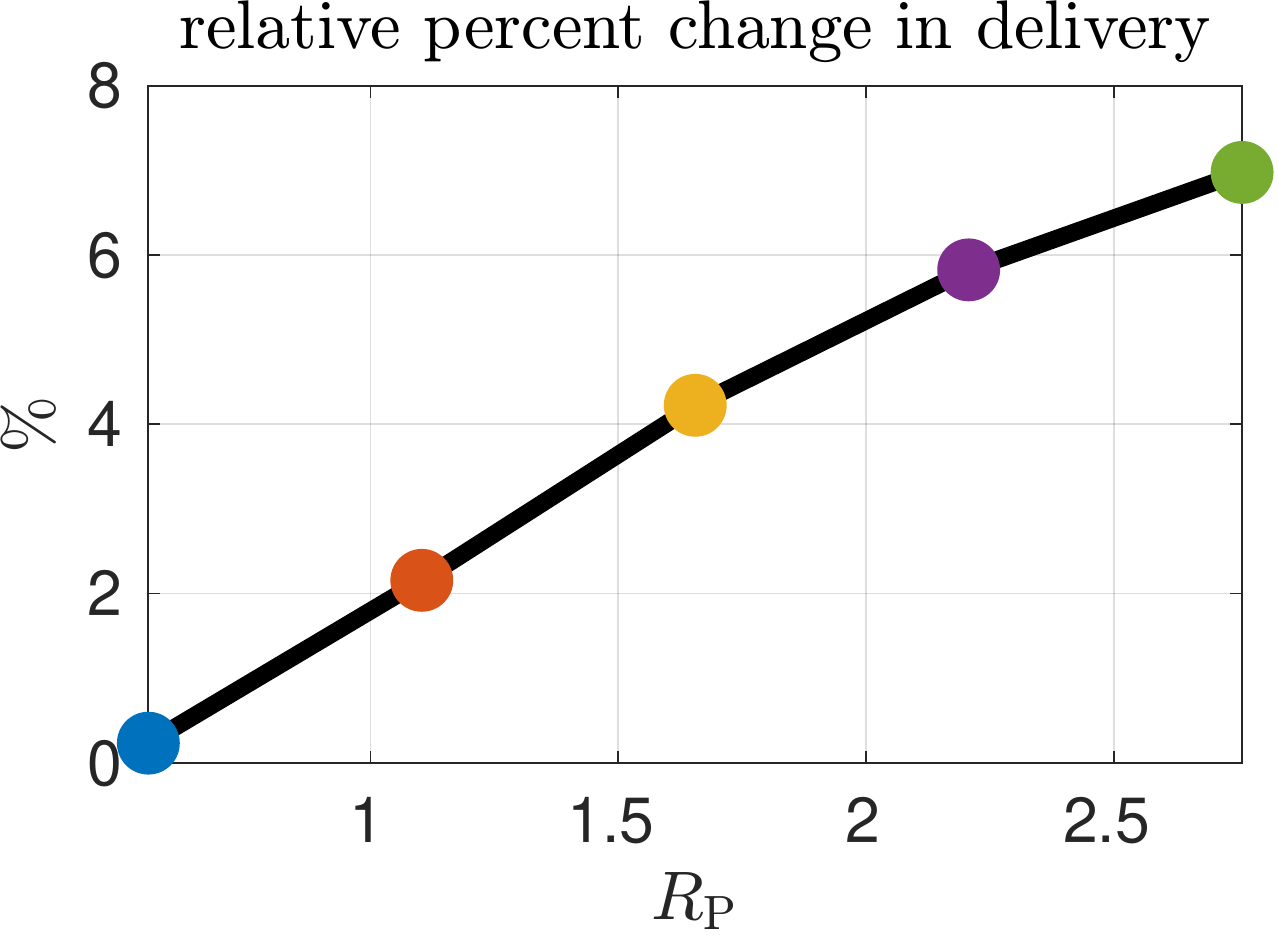}
\caption{Results corresponding to oxygen consumption equal to -0.154552 L min$^{-1}$. The left panel shows a zoomed in version of the oxygen delivery curves seen in panel (f) of Figure \ref{fig:results2}. Four of the five pulmonary vascular resistance values are displayed, and the optimal fenestration diameter for a given pulmonary vascular resistance is marked with a vertical dotted line. The right panel shows the relative change in delivery from the closed fenestration to the optimal fenestration as a function of the pulmonary vascular resistance.}
\label{fig:results3}
\end{center}
\end{figure}

In Figure \ref{fig:results3}, we plot the delivery curves for the four larger pulmonary vascular resistance values as well as the relative change in delivery (from closed fenestration to optimal fenestration) to study in greater detail the effect of pulmonary vascular resistance on the optimal fenestration diameter. The left panel of Figure \ref{fig:results3} is a zoomed-in version of panel (f) in  Figure \ref{fig:results2}, with vertical dotted lines indicating the optimal fenestration diameter. The right panel shows relative changes in delivery for all five values of the pulmonary vascular resistance. The relative change, from closed fenestration to optimal fenestration, is very small for the baseline pulmonary vascular resistance, and it increases for larger resistance values. This trend indicates that as the pulmonary vascular resistance decreases, the delivery curves transition from non-monotonic to monotonically decreasing functions of the fenestration diameter. In other words, an optimal fenestration diameter exists only for large enough pulmonary vascular resistances, and the relative effect of the fenestration on oxygen delivery is greater for larger pulmonary vascular resistances. 

\section{Conclusions}
In this paper, we have developed a pulsatile compartmental model of the Fontan circulation that describes blood flow and oxygen transport. It also incorporates a fenestration between the systemic and pulmonary veins. The model was calibrated to clinical data with the fenestration closed in order to create a baseline set of parameters. Then, we the studied the impact of an open fenestration on several hemodynamic and oxygen transport variables. An open fenestration decreased the pulmonary artery pressure and increased cardiac output, with larger impacts on these variables seen for larger values of the pulmonary vascular resistance. For our baseline oxygen consumption value and for pulmonary vascular resistances close to the baseline value, oxygen delivery monotonically decreased as a function of the fenestration diameter.  For pulmonary vascular resistances that are typical of at-risk patients, however, we see an increase in oxygen delivery for small fenestration sizes until an optimal size is reached, and this increase in oxygen delivery, although modest, is accompanied by a decrease in systemic venous ($\approx$ pulmonary aterial) pressure.  Thus, the reason for optimizing oxygen delivery may not so much be to increase oxygen delivery per se, but rather to ensure that oxygen delivery is not reduced by an intervention with a different benefit --- the reduction of systemic venous pressure. 

\section{Acknowledgements}
This work was supported in part by the Research Training Group in Modeling and Simulation funded by the National Science Foundation via grant RTG/DMS-1646339.

\clearpage

\bibliographystyle{plain}
\bibliography{optimal_fontan}

\begin{thebibliography}{10}

\bibitem{Barnea94}
Ofer Barnea, Erle~H Austin, Barbara Richman, and William~P Santamore.
\newblock {Balancing the circulation: theoretic optimization of
  pulmonary/systemic flow ratio in hypoplastic left heart syndrome}.
\newblock {\em {Journal of the American College of Cardiology}},
  24(5):1376--1381, 1994.

\bibitem{Barracano22}
Rosaria Barracano, Assunta Merola, Flavia Fusco, Giancarlo Scognamiglio, and
  Berardo Sarubbi.
\newblock {Protein-losing enteropathy in Fontan circulation: Pathophysiology,
  outcome and treatment options of a complex condition}.
\newblock {\em {International Journal of Cardiology Congenital Heart Disease}},
  page 100322, 2022.

\bibitem{Conover18}
Timothy Conover, Anthony~M Hlavacek, Francesco Migliavacca, Ethan Kung, Adam
  Dorfman, Richard~S Figliola, Tain-Yen Hsia, Andrew Taylor, Sachin
  Khambadkone, Silvia Schievano, et~al.
\newblock An interactive simulation tool for patient-specific clinical decision
  support in single-ventricle physiology.
\newblock {\em The Journal of Thoracic and Cardiovascular Surgery},
  155(2):712--721, 2018.

\bibitem{Degroff08}
CG~DeGroff.
\newblock {Modeling the Fontan circulation: where we are and where we need to
  go}.
\newblock {\em {Pediatric Cardiology}}, 29(1):3--12, 2008.

\bibitem{Egbe17}
Alexander~C Egbe, Heidi~M Connolly, William~R Miranda, Naser~M Ammash, Donald~J
  Hagler, Gruschen~R Veldtman, and Barry~A Borlaug.
\newblock {Hemodynamics of Fontan failure: the role of pulmonary vascular
  disease}.
\newblock {\em {Circulation: Heart Failure}}, 10(12):e004515, 2017.

\bibitem{Feldt96}
Robert~H Feldt, David~J Driscoll, Kenneth~P Offord, Ruth~H Cha, Jean Perrault,
  Hartzell~V Schaff, Francisco~J Puga, and Gordon~K Danielson.
\newblock {Protein-losing enteropathy after the Fontan operation}.
\newblock {\em {The Journal of Thoracic and Cardiovascular Surgery}},
  112(3):672--680, 1996.

\bibitem{Fontan71}
F~Fontan and E~Baudet.
\newblock Surgical repair of tricuspid atresia.
\newblock {\em Thorax}, 26(3):240--248, 1971.

\bibitem{Grosse13}
Lars Grosse-Wortmann, Andreea Dragulescu, Christian Drolet, Rajiv Chaturvedi,
  Yasuhiro Kotani, Luc Mertens, Katherine Taylor, Gustavo La~Rotta, Glen
  Van~Arsdell, Andrew Redington, et~al.
\newblock Determinants and clinical significance of flow via the fenestration
  in the fontan pathway: a multimodality study.
\newblock {\em International journal of cardiology}, 168(2):811--817, 2013.

\bibitem{Han21}
Seong~Woo Han, Charles Puelz, Craig~G. Rusin, Daniel~J. Penny, Ryan Coleman,
  and Charles~S. Peskin.
\newblock Computer simulation of surgical interventions for the treatment of
  refractory pulmonary hypertension, 2021.

\bibitem{Hijazi92}
Ziyad~M Hijazi, John~T Fahey, Charles~S Kleinman, Gary~S Kopf, and William~E
  Hellenbrand.
\newblock {Hemodynamic evaluation before and after closure of fenestrated
  Fontan. An acute study of changes in oxygen delivery.}
\newblock {\em Circulation}, 86(1):196--202, 1992.

\bibitem{Hoppensteadt13}
Frank~C Hoppensteadt and Charles~S Peskin.
\newblock {\em {Mathematics in Medicine and the Life Sciences}}, volume~10.
\newblock Springer Science \& Business Media, 2013.

\bibitem{Lemler02}
Matthew~S Lemler, William~A Scott, Steven~R Leonard, Daniel Stromberg, and
  Claudio Ramaciotti.
\newblock {Fenestration improves clinical outcome of the Fontan procedure: a
  prospective, randomized study}.
\newblock {\em Circulation}, 105(2):207--212, 2002.

\bibitem{Liang14}
Fuyou Liang, Hideaki Senzaki, Clara Kurishima, Koichi Sughimoto, Ryo Inuzuka,
  and Hao Liu.
\newblock {Hemodynamic performance of the Fontan circulation compared with a
  normal biventricular circulation: a computational model study}.
\newblock {\em {American Journal of Physiology-Heart and Circulatory
  Physiology}}, 307(7):H1056--H1072, 2014.

\bibitem{Marsden09}
Alison~L Marsden, Adam~J Bernstein, V~Mohan Reddy, Shawn~C Shadden, Ryan~L
  Spilker, Frandics~P Chan, Charles~A Taylor, and Jeffrey~A Feinstein.
\newblock {Evaluation of a novel Y-shaped extracardiac Fontan baffle using
  computational fluid dynamics}.
\newblock {\em {The Journal of Thoracic and Cardiovascular Surgery}},
  137(2):394--403, 2009.

\bibitem{Mondesert13}
Blandine Mond{\'e}sert, Fran{\c{c}}ois Marcotte, Fran{\c{c}}ois-Pierre Mongeon,
  Annie Dore, Lise-Andr{\'e}e Mercier, Reda Ibrahim, Anita Asgar, Joaquim Miro,
  Nancy Poirier, and Paul Khairy.
\newblock {Fontan circulation: success or failure?}
\newblock {\em {Canadian Journal of Cardiology}}, 29(7):811--820, 2013.

\bibitem{Mynard12}
JP~Mynard, MR~Davidson, DJ~Penny, and JJ~Smolich.
\newblock A simple, versatile valve model for use in lumped parameter and
  one-dimensional cardiovascular models.
\newblock {\em {International Journal for Numerical Methods in Biomedical
  Engineering}}, 28(6-7):626--641, 2012.

\bibitem{Ohuchi17}
Hideo Ohuchi.
\newblock {Where is the “optimal” Fontan hemodynamics?}
\newblock {\em {Korean Circulation Journal}}, 47(6):842--857, 2017.

\bibitem{Peskin86}
Charles~S Peskin and Cheng Tu.
\newblock Hemodynamics in congenital heart disease.
\newblock {\em Computers in Biology and Medicine}, 16(5):331--359, 1986.

\bibitem{Puelz17}
Charles Puelz, Sebasti{\'a}n Acosta, B{\'e}atrice Rivi{\`e}re, Daniel~J Penny,
  Ken~M Brady, and Craig~G Rusin.
\newblock {A computational study of the Fontan circulation with fenestration or
  hepatic vein exclusion}.
\newblock {\em {Computers in Biology and Medicine}}, 89:405--418, 2017.

\bibitem{Rychik97}
Jack Rychik, Jonathan~J Rome, and Marshall~L Jacobs.
\newblock {Late surgical fenestration for complications after the Fontan
  operation}.
\newblock {\em Circulation}, 96(1):33--36, 1997.

\bibitem{Rychik12}
Jack Rychik, Gruschen Veldtman, Elizabeth Rand, Pierre Russo, Jonathan~J Rome,
  Karen Krok, David~J Goldberg, Anne~Marie Cahill, and Rebecca~G Wells.
\newblock {The precarious state of the liver after a Fontan operation: summary
  of a multidisciplinary symposium}.
\newblock {\em {Pediatric Cardiology}}, 33(7):1001--1012, 2012.

\bibitem{Tu89}
Cheng Tu and Charles~S Peskin.
\newblock Hemodynamics in transposition of the great arteries with comparison
  to ventricular septal defect.
\newblock {\em {Computers in Biology and Medicine}}, 19(2):95--128, 1989.

\bibitem{Yang10}
Weiguang Yang, Jeffrey~A Feinstein, and Alison~L Marsden.
\newblock {Constrained optimization of an idealized Y-shaped baffle for the
  Fontan surgery at rest and exercise}.
\newblock {\em {Computer Methods in Applied Mechanics and Engineering}},
  199(33-36):2135--2149, 2010.

\end{thebibliography}

\end{document}